\documentclass{article}  

\newif\ifamsart
\amsartfalse 
\newif\ifieee
\ieeefalse 
\newif\ifllncs
\llncsfalse 
\newif\ifacm
\acmfalse 
\newif\ifrotatedtables
\rotatedtablestrue 

\usepackage{graphicx}
\usepackage{tikz}
\usetikzlibrary{arrows, positioning, shapes.geometric}
\usepackage{pdflscape}
\usepackage{verbatim}
\ifacm{}\else\usepackage{amssymb}\fi 
\usepackage{amsfonts}
\usepackage{amsmath}
\usepackage[shortlabels]{enumitem}
\usepackage{cite}
\usepackage[numbers,sectionbib]{natbib}
\usepackage{wrapfig}
\ifacm{
    \theoremstyle{definition}
    
    \newtheorem{definition}{Definition}[section]
  }\else
\fi
\usepackage{orcidlink}
\usepackage{booktabs}   
\usepackage{tabularx}   
\usepackage{caption}
\usepackage{subcaption}

\ifllncs\usepackage{smallsubsub}\else
\fi

\newcounter{running}[section]
\newenvironment{renumerate}{\begin{enumerate}\setcounter{enumi}{\value{running}}}%
{\setcounter{running}{\value{enumi}}\end{enumerate}}

\ifllncs
  \newtheorem{maxim}{Maxim}{\bfseries}{\upshape}

  \newtheorem{recomm}{Recommendation}{\bfseries}{\upshape}

\else\ifieee
    \newcounter{maximctr}[section]

    \newenvironment{maxim}{\par\refstepcounter{maximctr}\smallskip\noindent
      \textbf{Maxim \themaximctr.}  \em}{\hfill // \par\smallskip}
    
    \newtheorem{recomm}{Recommendation}
  \else
    \newtheorem{maxim}{Maxim}

    \newtheorem{recomm}{Recommendation}
  \fi
\fi

\newcommand{\super}[1]{\ensuremath{{}^{#1}}}

\newcommand{\constraininteraction}{Constrain possible interactions so
  attested properties depend only on limited, predictable, measurable
  parts of the system.}

\newcommand{\shortlived}{Long-lived services handling untrusted
  inputs need runtime remeasurement to recheck their state 
  invariants.  Short-lived, single-input processes
  avoid the need for runtime remeasurement.}

\newcommand{\resisttransitorycorruption}{Attestation should be
  independent of secrets that could be disclosed by transient
  corruptions, lest transient corruptions cause persistent attestation
  failure.}

\newcommand{\displayboottrait}{Each attestation must display
  traits showing that it originated from our
  uncorrupted
  attestation software, running under a trustworthy system boot.}

\newcommand{\ascenddependencies}{An acyclic dependency relation
  enables us to measure lower layers before components depending on
  them. Only very recent corruptions of underlying components can then
  undermine attestations.}

\ifieee
       
  \renewcommand\paragraph[1]{\par\medskip\noindent\textbf{#1}\hspace{1ex}}
  \newcommand\subparagraph[1]{\par\smallskip\noindent\textit{#1}\hspace{1ex}}
\fi

\ifamsart
\else\usepackage
  {hyperref}
\fi 

\usepackage{alltt}
\usepackage{array}
\usepackage{xcolor}
\usepackage{pifont}

\usepackage{xcolor}
\usepackage{listings}
\definecolor{dkgreen}{rgb}{0,0.6,0}
\definecolor{ltblue}{rgb}{0,0.4,0.4}
\definecolor{dkviolet}{rgb}{0.3,0,0.5}
\definecolor{dkblue}{rgb}{0,0,0.6}

\lstdefinelanguage{Coq}{ 
    mathescape=true,
    texcl=false, 
    morekeywords=[1]{Section, Module, End, Require, Import, Export,
        Variable, Variables, Parameter, Parameters, Axiom, Hypothesis,
        Hypotheses, Notation, Local, Tactic, Reserved, Scope, Open, Close,
        Bind, Delimit, Definition, Let, Ltac, Fixpoint, CoFixpoint, Add,
        Morphism, Relation, Implicit, Arguments, Unset, Contextual,
        Strict, Prenex, Implicits, Inductive, CoInductive, Record,
        Structure, Canonical, Coercion, Context, Class, Global, Instance,
        Program, Infix, Theorem, Lemma, Corollary, Proposition, Fact,
        Remark, Example, Proof, Goal, Save, Qed, Defined, Hint, Resolve,
        Rewrite, View, Search, Show, Print, Printing, All, Eval, Check,
        Projections, inside, outside, Def},
    morekeywords=[2]{forall, exists, exists2, fun, fix, cofix, struct,
        match, with, end, as, in, return, let, if, is, then, else, for, of,
        nosimpl, when},
    morekeywords=[3]{Type, Prop, Set, true, false, option},
    morekeywords=[4]{pose, set, move, case, elim, apply, clear, hnf,
        intro, intros, generalize, rename, pattern, after, destruct,
        induction, using, refine, inversion, injection, rewrite, congr,
        unlock, compute, ring, field, fourier, replace, fold, unfold,
        change, cutrewrite, simpl, have, suff, wlog, suffices, without,
        loss, nat_norm, assert, cut, trivial, revert, bool_congr, nat_congr,
        symmetry, transitivity, auto, split, left, right, autorewrite},
    morekeywords=[5]{by, done, exact, reflexivity, tauto, romega, omega,
        assumption, solve, contradiction, discriminate},
    morekeywords=[6]{do, last, first, try, idtac, repeat},
    morecomment=[s]{(*}{*)},
    showstringspaces=false,
    morestring=[b]",
    morestring=[d],
    tabsize=3,
    extendedchars=false,
    sensitive=true,
    breaklines=false,
    basicstyle=\small,
    captionpos=b,
    columns=[l]flexible,
    identifierstyle={\ttfamily\color{black}},
    keywordstyle=[1]{\ttfamily\color{dkviolet}},
    keywordstyle=[2]{\ttfamily\color{dkgreen}},
    keywordstyle=[3]{\ttfamily\color{ltblue}},
    keywordstyle=[4]{\ttfamily\color{dkblue}},
    keywordstyle=[5]{\ttfamily\color{dkred}},
    stringstyle=\ttfamily,
    commentstyle={\ttfamily\color{dkgreen}},
    literate=
    {\\forall}{{\color{dkgreen}{$\forall\;$}}}1
    {\\exists}{{$\exists\;$}}1
    {<-}{{$\leftarrow\;$}}1
    {=>}{{$\Rightarrow\;$}}1
    {==}{{\code{==}\;}}1
    {==>}{{\code{==>}\;}}1
    {->}{{$\rightarrow\;$}}1
    {<->}{{$\leftrightarrow\;$}}1
    {<==}{{$\leq\;$}}1
    {\#}{{$^\star$}}1 
    {\\o}{{$\circ\;$}}1 
    {\@}{{$\cdot$}}1 
    {\/\\}{{$\wedge\;$}}1
    {\\\/}{{$\vee\;$}}1
    {++}{{\code{++}}}1
    {~}{{\ }}1
    {\@\@}{{$@$}}1
    {\\mapsto}{{$\mapsto\;$}}1
    {\\hline}{{\rule{\linewidth}{0.5pt}}}1
  }[keywords,comments,strings]

\ifacm{
    \setcopyright{acmlicensed}
    \copyrightyear{2025}
    \acmYear{2025}
    \acmDOI{XXXXXXX.XXXXXXX}
    \acmConference[Conference Acronym '25]{Conference Full Name}{Date 1--31,
      2025}{Location}
    \acmISBN{XXX-X-XXXX-XXXX-X/2025/XX}
  }
\fi

\newcommand{\hash}[1]{\ensuremath{\mathsf{hash}(#1)}}

\newcommand{\pcraction}[4]{#1 & #2 & #3 & #4 }
\ifieee
  \newcommand{\resizesf}[1]{{\small{\textsf{#1}}}}
\else
  \newcommand{\resizesf}[1]{{\textsf{#1}}}
\fi
\newcommand{\ask}{\ensuremath{\mathop{ASK}}}
\newcommand{\uefi}{\texttt{uefi}}
\newcommand{\grub}{\texttt{GRUB}}
\newcommand{\maestro}{Maestro}
\newcommand{\attestationattuned}{attestation-attuned}

\begin{document}
\title{Designing Trustworthy Layered Attestations
}
%
%

\ifllncs\authorrunning{Thomas, Schmalz, Petz et al.}\fi
\ifieee{
    \author{\IEEEauthorblockN{Will Thomas, Logan Schmalz,}
      \IEEEauthorblockN{Adam Petz, Perry Alexander}
      \IEEEauthorblockA{University of Kansas\\ Institute for Information Sciences \\
        \{30wthomas,loganschmalz,ampetz,palexand\}@ku.edu}
      \and
      \IEEEauthorblockN{Joshua D. Guttman, Paul D. Rowe}
      \IEEEauthorblockA{The MITRE Corporation\\
        \{guttman,prowe\}@mitre.org}
      \and
      \IEEEauthorblockN{James Carter}
      \IEEEauthorblockA{National Security Agency\\
        jwcart2@uwe.nsa.gov}}}
\else{
    \author{ {Will Thomas}\super{1} \and {Logan Schmalz}\super{1} \and
      {Adam Petz}\super{1} \and 
      {Perry Alexander}\super{1} \and {Joshua D. Guttman} \and
      {Paul D. Rowe}\super{2} \and 
      {James Carter}\super{3}}
    
    \ifllncs{ \institute{{The University of Kansas}  \\
          {The MITRE Corporation}  \\
          {National Security Agency}}} %
    \fi}
\fi

\ifllncs{}\else\ifacm{}\else\date{\normalsize \super{1}The University of Kansas
      \quad \super{2}The MITRE Corporation \\ \super{3}National Security Agency}\fi\fi

%

%
%


\maketitle
\ifacm\else{
    \begin{abstract}

      %
\emph{Attestation} means providing evidence that a remote \emph{target
  system} is worthy of trust for some sensitive interaction.  Although
attestation is already used in network access control, security
management, and trusted execution environments, it mainly concerns
only a few system components.  A clever adversary might manipulate
these shallow attestations to mislead the relying party.

Reliable attestations require \emph{layering}.  We construct
attestations whose layers report evidence about successive components
of the target system.  Reliability also requires structuring the
target system so only a limited set of components matters.


We show how to structure an example system for reliable attestations
despite a well-defined, relatively strong adversary.  It is based on
widely available hardware, such as Trusted Platform Modules, and
software, such as Linux with SELinux.  We isolate our principles in a
few maxims that guide system development.  We provide a cogent
analysis of our mechanisms against our adversary model, as well as an
empirical appraisal of the resulting system.
%
%
We also identify two improvements to the mechanisms so attestation can
succeed against strengthened adversaries.  The performance burden of
our attestation is negligible, circa~1.3\%.

After our first example, we vary our application level, and then also
its underlying hardware anchor to use confidential computing with
AMD's SEV-SNP.  The same maxims help us achieve trustworthy
attestations.



    \end{abstract}
  }
\fi 

\pagestyle{plain}

%
%
%

%

\section{Introduction}
\label{sec:intro}


Good security mechanisms can prevent bad things from happening and
reduce the harms when bad things do happen. Good attestation
mechanisms play a complementary role: though they don’t prevent bad
things from happening, they make sure the bad news gets to those who
need to hear it. Poor security mechanisms can make reliable
attestation hard by interfering with undermining the attestation
mechanisms. In this paper we aim to determine ways of building good
attestation mechanisms, ensuring that they have the security
infrastructure they need to be effective, and determining what bad
things they will reliably report.

Attestation is about trust (and distrust).  To deliver sensitive
information to a system, we need to trust the system will not
mishandle the information contrary to our intent.  When acting on
information from a system, we rely on it not to mislead us.  Even if
the system was carefully designed with security mechanisms to prevent
such misbehavior, we know no security design is immune to malicious
attacks.  Malware can gain root user privileges, and can even hide
itself from security tools.  It can hijack core mechanisms embedded in
the operating system kernel.  There are, therefore, reasons to
distrust even well-designed systems.

Attestation \emph{discloses} rather than \emph{defeats} adversary
attacks that make the system untrustworthy.  The purpose of an
attestation is to provide the \emph{relying party}---who will engage
in a \emph{sensitive operation} with the system---with \emph{evidence}
to justify trust in the current state of the system.  That is, a
skeptical \emph{appraisal} of the evidence will hopefully justify the
inference that no adversary \emph{currently} has a foothold on the
system that would allow it to interfere with the proper processing of
our information.  Hence, an attestation should inspect, or \emph{measure},
aspects of the \emph{runtime} state.  The \emph{appraiser} evaluating
the evidence may be the relying party or an expert already trusted by
the relying party for this purpose.

Runtime attestation poses a big problem for a cautious appraiser:
Trusting the evidence in an attestation requires a trustworthy
attestation infrastructure on the system itself to produce it.  How
can we gain trust in the attestation infrastructure itself?  Appealing
only to runtime evidence never grounds our trust in anything solid.
Instead, we need a reliable root of trust and a way to bootstrap trust
from the root to other system components.  When the adversary has
software-only access, hardware mechanisms can provide the reliability
we need.  This suggests that runtime attestation needs to be
\emph{layered}, combining mechanisms from at least two layers
(hardware and software).  In fact, well designed systems include
security mechanisms that allow the software to be further subdivided
into more layers with differing levels of exposure to an adversary,
yielding differing levels of skepticism about their susceptibility to
attack.

Trustworthy attestation thus requires a delicate composition of
security and attestation mechanisms at various system layers that must
all cooperate (in subtle ways, as we will see) to generate a body of
evidence robust enough to justify an appraiser's inference that the
system is currently in a trustworthy state.

Although attestation and appraisal are already in use in many
situations
\cite{nunes2019vrased,carpent2018erasmus,Brickell:04:Direct-anonymou,
  Gu:2008:Remote-Attestation-on-Program-Execution,carpent2017lightweight,
  sultanasy22acase}, combining existing mechanisms in effective ways
is difficult. Some mechanisms on their own are \emph{shallow},
providing only a few types of information, for example, targeting
userspace malware without addressing more advanced adversaries that
could install a rootkit to hijack kernel mechanisms to avoid
detection. Other mechanisms are \emph{untimely}, in recording
information only at boot, failing to detect runtime attacks that are
possible, especially if the system may run for a long time between
boot cycles. Moreover, attempts to be comprehensive and timely may end
up being \emph{fragile} by delivering, say, one opaque value to
summarize the system state, making it unmanageable to appraise trust
as system components are updated piecemeal.

The attestation system plays a game with the adversary.  The
adversary's moves may corrupt or replace system components, and also
may restore or repair components so attestations will succeed.  The
attestation system's moves include taking measurements and combining
results in a protocol-compliant order.

The adversary wins a run of the game if appraisal deems the system
compliant while the system is in fact corrupted.  Otherwise the
attestation system wins.

An attestation system that never appraises a system as compliant
always wins, as does any attestation system running on an
incorruptible system.  Naturally, we aim for attestation systems that
tell us more than the first and apply to reality better than the
second.  The adversary is another variable.  An unrealistically strong
adversary may have a winning strategy, although the attacks actual
adversaries can deliver may not yield winning strategies.  Thus, in
designing the security and attestation mechanisms, we should also
define the class of adversaries against which our design can succeed.

This paper has two primary goals. First we aim to show that it is
possible to design effective combinations of existing attestation and
security mechanisms to achieve trustworthy layered runtime
attestation. Secondly, we aim to share with the reader some mental
tools we have identified that have helped us think critically about
how to design and evaluate layered attestation systems.

In this paper we focus on a concrete example that helps to ensure we
are grounded in realistic concerns.  We use very standard hardware,
such as Trusted Platform Modules, and software, such as commodity
Linux with Security Enhanced Linux~\cite{Loscocco2001b} and the
Integrity Measurement Architecture~\cite{Sailer:04:Design-and-impl}.
The example application is a Cross-Domain Solution (CDS) that sits at
the boundary between networks with different security postures, and
rewrites and filters messages of specific kinds before they cross the
boundary.  We fully implemented security and attestation mechanisms
for a CDS.  Our work illustrates how to use adversarial reasoning to
design the mechanisms and---in tandem---to identify a clear
\emph{adversary model} against which the attestations can succeed.
The general considerations that drove the process are formulated in a
set of five \emph{design maxims} that help thinking critically about
layered attestation systems.

To show they are helpful beyond one example, we consider a different
application level context as well as a different hardware root of
trust, namely confidential computing with a Trusted Execution
Environment rather than the TPM.  Our strategy in each of these two
variants revolves around the same maxims.

\paragraph{Contributions.}  We make five contributions:
\begin{enumerate}
  \item\label{item:contribution:design} We identify an adversary model
  capturing advanced and realistic threats against Linux system
  equipped with Security-Enhanced Linux and Linux Integrity
  Measurement Architecture.  The adversary model was a byproduct of
  how we designed and evaluated this layered attestation system.
  \item\label{item:contribution:winning} A combination of empirical
  and formal evidence shows that our mechanisms win the attestation
  game by defending against these threats or detecting successful
  corruptions.
  \item\label{item:contribution:maxims} Five reusable maxims summarize
  the adversary-driven reasoning that led to our solution.
  \item\label{item:contribution:variants} Two major variants of our
  system show how the same maxims help shape attestation solutions for
  them.
  \item\label{item:contribution:recommendations} We make two
  recommendations to strengthen existing kernels and architectures to
  eliminate two shortcomings we encountered in how we followed our own
  maxims.  The very standard mechanisms we worked with imposed these
  shortcomings on us.  We limited our adversary to rule out the
  (challenging) attacks that exploit these shortcomings.
\end{enumerate}

We consider it an advantage that our maxims allow us to identify
compromises that existing hardware, software, and performance may
impose on us, and that they aid in identifying specific improvements
as in contribution~(\ref{item:contribution:recommendations}).

Section~\ref{sec:downward} is devoted to
contributions~(\ref{item:contribution:design})
and~(\ref{item:contribution:maxims}), with Section~\ref{sec:build:up}
adding detail to the CDS design.  Section~\ref{sec:build:up:lifetime}
discusses recommendation~\ref{recomm:lkim:acyclic},
and~\ref{sec:build:up:evidence} discusses
recommendation~\ref{recomm:tpm:localities}.
Section~\ref{sec:self:test} handles the
evaluation~(\ref{item:contribution:winning}), while
Section~\ref{sec:use-cases} moves away from the CDS example to apply
the maxims to the variant
applications~(\ref{item:contribution:variants}).  %

\section{Mechanisms and Standards}
\label{sec:mechansisms:standards}

We start from a selection of widely available mechanisms for security
and attestation.
%
%
Trustworthy layered attestation requires judiciously combining
existing technologies; success requires craftsmanship.
%
%
In this section, we briefly summarize
%
%
several core mechanisms and standards for attestation so that in
subsequent sections it will be clear how each part contributes to
create a trustworthy whole. Our choices do not aim to be
comprehensive.  They focus on technologies that figure into the
designs we will explore.

\paragraph{Summary of our candidates.}
Reliable attestation without hardware support seems hopeless: A remote
appraiser would receive the same attestation if a lying OS controlled
the
processor~\cite{ElDefrawy:2012:SMART-Secure-and-Minimal-Architecture-for-Establishing-Dynamic-Root-of-Trust,
  Castelluccia:2009:Difficulty-of-Software-Based-Attestation}.
\nocite{FarmerEtAl1996}
The past decades have seen hardware support develop, starting with the
\emph{Trusted Platform
  Modules}~\cite{Trusted-Computing-Group:2025:TCG-TPM-2.0-Part0-Introduction},
followed by a number of Trusted Execution Environments (TEEs) for
\emph{confidential computing}.  Intel's Software Guard Extensions
SGX~\cite{cryptoeprint:2016:086} were not a basis for attesting to
other portions of the system, since an SGX enclave is always a thread
within a user-level process, and prohibited from executing system
calls.  However, AMD's
SEV~\cite{AMD:2020:SEV-SNP-VM-Isolation-Whitepaper} and Intel's more
recent TDX both offer a virtual machine as the unit of protection.
Other technologies exist at the kernel layer. These include
\emph{Security-Enhanced Linux}~\cite{Loscocco2001b}, which implements
the Flask security architecture~\cite{Loscocco2001a},
cf.~\cite{Guttman:04:Verifying-infor,Hicks:07:A-logical-speci}, as
well as the Linux \emph{Integrity Measurement Architecture}
IMA~\cite{Sailer:04:Design-and-impl} and \emph{Linux Kernel Integrity
  Measurer}
LKIM~\cite{Loscocco:07:Linux-kernel-in,thober2008improving} which
perform measurements of applications or indeed of the kernel itself.
At the application layer, \emph{Copland} is a language for attestation
protocols~\cite{Ramsdell:2019aa} and \emph{{\maestro}} implements
it~\cite{Petz:2024:Verified-Configuration-and-Deployment-Paper}.
RATS offers \emph{Remote Attestation standards}~\cite{ietf2023rats}
useful for interoperability among the various technologies.
The remainder of section gives more detail about each of the above.

\paragraph{Trusted Platform Modules.}  A Trusted Platform Module is a
small bit of hardware functionality; while originally a separate
device connected to the CPU by a bus, it is now increasingly
incorporated within the CPU itself.
%
%
The TPM has two relevant
capabilities.  One is to store cryptographic keys such as signing
keys, using them to generate signatures under specified constraints.
It can emit these keys in a wrapped (i.e.~encrypted) form, so the key
can be used only when loaded back into the TPM under the same usage
constraints.  Thus, it serves as a \emph{root of trust for storage} of
values like these keys.

Second, a TPM has a collection of \emph{Platform Configuration
  Registers} (PCRs) whose values contribute to the usage constraints
on keys.  Some of the PCRs are reset only when the platform is powered
down, and are otherwise modified only by having their values
\emph{extended}.  To extend the value of a PCR, the TPM hashes its old
value $v$ together with a newly presented value $w$, placing the
result in the PCR as its new value $u=\hash{\mathsf{concat}(v,w)}$.

If the platform's boot-time hardware (e.g.~{\uefi}) is designed to
extend a PCR with the hash of the executable image to which it will
pass control, and that executable acts similarly when passing control
to new code, then this PCR can build up a hash chain that
unambiguously reflects the sequence of images controlling the platform
in its early boot phases.  Since this information is reflected in the
digital signatures the TPM will generate, a remote party can check
them to identify the software in control in the early phases.  This is
a core hardware basis for trustworthy remote attestation, or a
\emph{root of trust for reporting}.  We use the TPM in our main case
study.

\paragraph{Confidential Computing.}  A more recent alternative is to
use TEEs like AMD's SEV-SNP or Intel TDX as a basis for attestation.
Each of them launches a virtual machine in a trusted memory space,
ensuring that its memory cannot be read by the underlying hypervisor.
A processor-authenticated record can associate the hash of the initial
image controlling this VM with a signature-verification key; the
corresponding signing key may be maintained as a secret within the VM.
We will discuss confidential computing as an alternative basis for
layered attestation in Section~\ref{sec:amd_use-case}.

\paragraph{Security-Enhanced Linux.}  The Linux Security Module
framework gives a module security hooks at strategic points in the
kernel, and SELinux is a particular module that plugs into it.
SELinux associates a \emph{security context} with each active process
and with many system resources such as files, sockets, etc.  A
\emph{policy} supplies rules that may prevent processes from acting on
resources according to their security contexts.  The security context
of an executable file also constrains the context of a process that
\texttt{exec}s it; when a process forks, the parent process's context
constrains that of the child.  This allows an SELinux policy to
control the transitive effects of processes as they launch children
and act on system resources.  Optionally, the SELinux configuration
can be made immutable at runtime.  This is a frequent choice, which we
will follow.  Thus, the SELinux configuration chosen at boot time will
be in force throughout the runtime of that boot cycle, unless there
has been a kernel corruption.

\paragraph{Integrity Measurement Architecture.}  Technologies for runtime measurement
are not nearly as widespread as the technologies described above. 
Like SELinux, the Linux Integrity Measurement Architecture
(IMA)
is also a module loaded into the Linux kernel. The IMA policy 
selects files in the file system, e.g.~via their file owner or SELinux
security contexts.
When one of these files is accessed for a selected operation
(e.g. read, write, exec), IMA hashes its contents.
This hash value may be extended into a TPM
register; in addition, IMA may raise an error if the computed hash
value differs from its required hash value.  We use IMA in this latter
mode to ensure that when our sensitive processes launch, the running
executable code will match our requirement.
The IMA policy can only be appended at runtime, and later
rules cannot override earlier ones; therefore, it is effectively immutable.

\paragraph{Linux Kernel Integrity Measurement}
In contrast to IMA, the Linux Kernel Integrity Measurement (LKIM)
tool~\cite{Loscocco:07:Linux-kernel-in} is designed for the purpose of
re-measuring the state of the \emph{running} Linux kernel.  It is the
result of careful design to uncover the vast bulk of modifications
that an adversary could use to maintain control in a Linux kernel.
LKIM rechecks core control flow and data structure invariants of the
Linux kernel. Although the adversary's initial point of entry may be
some as-yet unimagined zero-day kernel flaw, benefiting from the flaw
requires a persistent alteration to the kernel's invariants and future
control flow.  There is a relatively limited set of candidates to
achieve persistence.  Much craftsmanship was needed to construct LKIM
(and a corresponding Windows remeasurement tool called WinKIM), but
the effort may be amortized over many kinds of attestation based on
the operating system.

\paragraph{Copland and its {\maestro} implementation.}  Layered
attestations will require a variety of steps to be taken, some in
specific orders, at one or more execution locations in a platform or
distributed system.  Copland is a language to specify layered
attestation protocols, i.e.,~which steps should be taken at which
locations, under which ordering constraints.  {\maestro} is a toolkit
that serves as a compiler for Copland, generating configurations and
executables to ensure that a Copland protocol can run to completion,
yielding evidence compatible with its semantics.

\paragraph{The IETF Remote Attestation Procedures.}  RATS defines the
fundamental roles of \textit{attester}, \textit{verifier}, and
\textit{relying party}, terminology that overlaps with ours but
differs slightly from our \emph{target of attestation} and
\emph{appraiser}.  RATS also standardizes the syntactic structure of
attestation evidence.  While {\maestro} does not currently generate its
evidence in the RATS-standardized form, it will become a fully
RATS-compliant tool.
%
%
RATS makes no attempt to express how to orchestrate attestations,
i.e.,~which combinations of evidence make sense, and which orderings on
evidence collection are reliable.  Hence, Copland and {\maestro} fill a
gap that RATS makes no attempt to cover.

\paragraph{The challenge.}  Our central question in this paper is how
to use the techniques described above.  They ensure that we can
generate and deliver pieces of evidence.  But none of them give a
routine path to the trustworthy attestation we need.  For instance, a
virtual machine running Linux inside a TDX or SEV TEE is protected
from other activities on its platform, but it is just as vulnerable to
data-driven attacks and rootkits as if Linux were handling the same
inputs in the same way natively
\cite{Niu:2026:What-You-Trust-is-Insecure}.  Hence we ask:  what
approaches will yield attestation data that are a trustworthy basis
for appraising our remote peers?


\section{Decomposing for Attestation}
\label{sec:downward}

In this section we consider a particular example system.  We use
adversarial reasoning to infer the types of security mechanisms it
needs, and to identify relevant attestation technologies suggested by
these needs. This suggests the outlines of a design, the details of
which we flesh out in Section~\ref{sec:build:up}. An outcome of this
process is an adversary model that clearly identifies the types of
adversary capabilities we aim to defend against.
While this section focuses on one target example, we identify along
the way five maxims for designing trustworthy layered attestations
that we contend have broad applicability. In
Section~\ref{sec:use-cases} we demonstrate how these maxims might be
applied to the design and evaluation of alternative design patterns
for different use cases.

\subsection{Example System:  A Cross-Domain Solution}
\label{sec:downward:example}

\begin{figure*}
  \centering{\footnotesize
    \begin{tikzpicture}[->,>=stealth',shorten >=1pt,auto,node distance=2.0cm,
  thick,main node/.style={rectangle,fill=blue!20,draw,
    font=\sffamily,minimum height=6mm,minimum width=10mm},
  buffer node/.style={rectangle,fill=green!30,draw,
    font=\sffamily,minimum height=6mm,minimum width=5mm},
  config node/.style={rectangle,fill=purple!60,draw,
    font=\sffamily,minimum height=5mm,minimum width=5mm},
  io node/.style={rectangle,
    font=\sffamily,minimum height=5mm,minimum width=10mm},
  remote node/.style={rectangle,fill=brown!20,draw,
    font=\sffamily,minimum height=6mm,minimum width=10mm}]

  \node[remote node] (high) {RN};
  \node[main node] (intake) [node distance=2cm, right of=high] {Intake};
  \node[buffer node] (incoming) [node distance=1.2cm, right of=intake] {};
  \node[main node] (rewrite) [node distance=1.2cm, right of=incoming] {Rewrite};
  \node[buffer node] (rewritten) [node distance=1.2cm, right of=rewrite] {};
  \node[main node] (filter) [node distance=1.2cm, right of=rewritten] {Filter};
  \node[buffer node] (outgoing) [node distance=1.2cm, right of=filter] {};
  \node[main node] (export) [node distance=1.2cm, right of=outgoing] {Export};
  \node[remote node] (low) [node distance=2cm, right of=export] {ON};

  \node[io node] (incomingLab) [node distance=0.6cm, above
  of=incoming]{Incoming};
  \node[io node] (rewrittenLab) [node distance=0.6cm, above
  of=rewritten]{Rewritten};
  \node[io node] (outgoingLab) [node distance=0.6cm, above
  of=outgoing]{Outgoing};

  \path[every node/.style={font=\sffamily\small, fill=white,inner sep=1pt}]
  (high) edge (intake)
  (intake) edge (incoming)
  (incoming) edge (rewrite)
  (rewrite) edge (rewritten)
  (rewritten) edge (filter)
  (filter) edge (outgoing)
  (outgoing) edge (export)
  (export) edge (low)
  ;
  \node[config node] (rewriteConfig) [node distance=1.25cm, below of=rewrite] {};
  \node[config node] (filterConfig) [node distance=1.25cm, below of=filter] {};
  \node[io node] (configLab) [node distance=1.25cm, below of=rewritten]{Config Files};

  \path[every node/.style={font=\sffamily\small, fill=white,inner sep=1pt}]
  (rewriteConfig) edge (rewrite)
  (filterConfig) edge (filter)
  ;


\end{tikzpicture}


}
  \caption{Cross-Domain System:  Messages from Restricted Network to
    Open Network}
  \label{fig:cds}
\end{figure*}
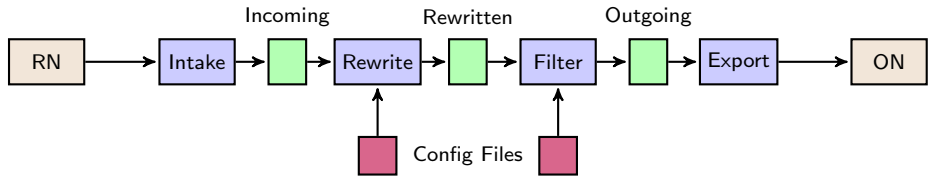

Suppose that we have two networks with different security postures,
and would like to deliver some messages, in a supervised way, from the
more restricted network \resizesf{RN} to the more open network
\resizesf{ON}.  For this we could use a Cross-Domain Solution (CDS),
meaning a computer with a network interface on each of the two
networks.  Fig.~\ref{fig:cds} illustrates the application structure.
An \resizesf{intake} server process listens on \resizesf{RN} for
connections, each of which will deliver a validly formatted email
message to the CDS.  The intake process makes the message available to
a \resizesf{rewrite} process, which may modify or sanitize some of its
headers to make it suitable for delivery, discarding irreparable
messages.  A \resizesf{filter} process receives the surviving
messages, and, without altering them, makes a final determination on
which are acceptable for delivery to \resizesf{ON}.  These messages
are delivered to an \resizesf{export} process, which connects through
a socket on the \resizesf{ON} interface to a server on \resizesf{ON}
for delivery.  The rewrite and filter processes are designed for a
good degree of generality, and each one has a configuration file that
determines the message processing rules it will apply.
The application processes use the directories \resizesf{Incoming},
\resizesf{Rewritten}, and \resizesf{Outgoing} to pass messages to
successive stages.
\subsection{Security Goals}
\label{sec:downward:goals}

The CDS is a good example for our purposes, since it provides a
straightforward security service to the networks.  Namely, any message
leaving by the \resizesf{export} should have been
\begin{itemize}
  \parskip=0pt\itemsep=0pt
  \item input from \resizesf{RN} via the \resizesf{intake} process
  \item rewritten by the \resizesf{rewrite} process using the authorized
        configuration file
  \item output by the \resizesf{filter} process only if no further
        rewrites would be required by its authorized configuration file
  \item exported only when received from the filter process.
\end{itemize}
Thus, this is a \emph{pipeline}:  Messages should enter the pipeline
only at the \resizesf{intake}; a message should be delivered only if
it reaches the \resizesf{export}; no message should skip any pipeline
stage; and each step is taken by a known process, controlled by an
authorized configuration file~\cite{Boebert1985}.

This must be the \emph{only} pipeline from \resizesf{RN} to
\resizesf{ON} on the platform; below we will see how SELinux can help
ensure that only \resizesf{intake} has access to the platform's
interface on \resizesf{RN} and only \resizesf{export} has access to
the interface on \resizesf{ON}.
%

\subsection{Application Threats}
\label{sec:downward:threats}

The description of the application and its goals immediately makes
some potential threats clear.

\paragraph{Pipeline threats.}
\label{sec:downward:threats:external}
Can the adversary, by controlling processes on the platform, read or
modify directories and files on our system?
\begin{enumerate}
  \item Messages may enter the \resizesf{Incoming},
        \resizesf{Rewritten}, and \resizesf{Outgoing} directories without
        traversing earlier processes.
  \item A configuration file may be modified to rewrite or filter a
        message via an unauthorized policy.
  \item An adversary may replace an executable file for a process,
        causing it to pass undesirable messages.
\end{enumerate}
There are two strategies to respond.  One is to engineer the whole
system so we can be confident the adversary can never read or modify
directories or files on our system.  This strategy is intrinsically
fragile~\cite{Loscocco:98:The-Inevitabili}.  When new software has
vulnerabilities accessible from the network, disaster may ensue.

The alternate strategy ensures that the properties we care about
depend only on a small set of programs and functionality, and that a
few robust mechanisms ensure other software on the system cannot
damage the outcomes we are seeking.  Applied to attestation, this
means that the trustworthiness of an operation should depend only on a
small number of mechanisms, each of which can be measured and the
results assembled into a self-contained attestation.

\begin{maxim}\label{maxim:constrain:interaction}
  \constraininteraction
\end{maxim}

A natural mechanism to leverage for this purpose is
SELinux~\cite{Mayer:07:SELinux-by-Exam}. With a judicious use of
security contexts applied to the pipeline directories, binaries, and
configuration files, SELinux can enforce the pipeline property.
The integrity of these executables can further be protected using
IMA~\cite{sailer2004design}.

Thus, we need a way to attest to the fact that SELinux and IMA are
running, and controlled by the intended policies.  We also need to
attest that the application-level processes and the underlying kernel
will behave according to their expected behavioral semantics, and are
not hijacked by the adversary.  We turn next to these two concerns,
taking the behavioral concern first, which is often undermined by
data-driven corruptions.

\paragraph{Data-driven corruption threats.}
\label{sec:downward:threats:corruption}
Crafted inputs cause unintended behavior in flawed programs.  Attacks
including stack-smashing and return-oriented
programming~\cite{Shacham:2007:Geometry-of-Innocent-Flesh-on-the-Bone}
deliver data to change a program's control flow from the semantics of
its source code.

In a CDS, a crafted message may cause a process to handle subsequent
messages wrongly.  An adversary could cause a succession of sensitive
messages to pass to the open network \resizesf{ON}, disclosing
contents that were not already under the control of the adversary.

Memory-safe programming languages including Haskell, OCaml, and
Rust~\cite{hutton2016programming,leroy25:ocaml,klabnik2023rust} resist
these attacks with a correct compiler.\footnote{We chose to write our
  application programs in OCaml.}
However, compilers may be imperfect, especially for security
properties~\cite{xu2023silent}, and applications often involve foreign
code in languages such as C; system software is often written
predominantly in C.

The effect of the maliciously crafted input is generally to falsify
the invariants that the compiler attempts to maintain.  For instance,
the call stack should reflect the continuations the source program
predicts in the remainder of the computation.  Stack-manipulating
attacks break this correspondence.

Two potential approaches could prevent persistent misbehaviors.  One
would inspect the memory of the long-running process periodically,
ensuring its control state is compatible with the control flow graph
of the source program, and that its data structures respect the
invariants expected for this control state. This is \emph{runtime
  process remeasurement}.  It is delicate and must be based on the
actual source code of the specific program.

The alternative requires much less craftsmanship: Start a separate
process to handle each message.  A highly crafted message may be able
to cause unintended behavior in the process that handles it, but does
not leave behind a corrupted process to misdeliver successive
messages.  They are handled by fresh invocations.  This suggests a
design in which the server forks a new \resizesf{intake} process for
each incoming connection from \resizesf{RN}; if it never actually
handles any message itself, it would thus be hard to degrade.  The
\resizesf{intake} process could then invoke a \resizesf{rewrite}
process for the one message it has received.  Similarly, instances of
the \resizesf{filter} and \resizesf{export} processes could be started
by their predecessor in the pipeline as the predecessor's last action.
This would avoid contagion from one message to later messages.

Not all cases can be handled in this second way.  The kernel itself
cannot easily be restarted regularly, and its address space can be the
target of persistent attacks, namely rootkits.  \emph{Kernel runtime
  remeasurement} is therefore called for.  LKIM, the Linux Kernel
Integrity Measurement tool~\cite{Loscocco:07:Linux-kernel-in}, is
designed precisely for this purpose.  
%
\begin{maxim}\label{maxim:short:lived}
  \shortlived
\end{maxim}
%

\subsection{Safeguarding Attestation Mechanisms}
\label{sec:downward:underlying}

Even if LKIM and the additional measurers determine that the platform
is currently compliant,
%
%
it is infeasible to determine whether the platform has \emph{always}
been compliant.  Over time it may be booted in different ways, with
different initialization programs.  Indeed, normal management
practices require evolution and corrections, meaning that the system's
critical resources such as its filesystem may also have been available
during untrustworthy boots.  This is one kind of \emph{transitory
  corruption}.

Transitory corruption may also occur if the adversary briefly inserts
a rootkit into the kernel, possibly removing the rootkit before kernel
runtime remeasurement with LKIM occurs.  Either of these forms of
corruption could give the adversary access to any secrets resident in
kernel memory or the user level processes.

Disclosing a secret is a persistent result even if the cause is
transitory.  Thus we should avoid relying on memory resident secrets:
\begin{maxim}\label{maxim:resist:transitory:corruption}
  \resisttransitorycorruption
\end{maxim}
Signing keys are a crucial case, as attestation evidence must travel
across a network to reach its appraiser.  The signing key must be used
by our attestation mechanism, but must be unavailable to other parts
of the system to remain unavailable to the adversary.
Maxim~\ref{maxim:resist:transitory:corruption} says we cannot store
this signing key in the kernel or in application-level processes
handling attestation.  A natural choice would be to maintain it in a
TPM instead. 
We call this special key the \emph{Attestation Signing Key} \ask{}.

A TPM can protect the confidentiality of \ask{} in the following way:
\begin{itemize}
  \item The attestation signing key \ask{} is TPM resident in the
        sense that it exists in unwrapped form only in the TPM;
  \item It is exported from the TPM only in wrapped form, encrypted
        together with a signing policy that controls when the TPM will use
        it to generate signatures;
\end{itemize}

Although a TPM can protect the \ask{} from disclosure due to transient
corruptions, this protection is not sufficient if an adversary can
request signatures using this key, allowing it to forge attestations.
This suggests another core principle.  An appraiser never has a
reason to accept purported evidence if that evidence could emerge from
a system that might misuse signing keys, or for that matter generate
the evidence to be signed by mere assertion rather than by actual
measurement.  Thus:
\begin{maxim}\label{maxim:display:boot:trait}
  \displayboottrait
\end{maxim}

TPMs offer a relevant mechanism that we could leverage. Namely:
\begin{itemize}
  \item The wrapped form of \ask{} exported by the TPM can also contain
        a signing policy that controls when the TPM will use it to generate
        signatures;
  \item The signing policy guarantees it can sign data only if TPM
        Platform Configuration Registers (PCRs) have values guaranteeing a
        desired outcome from the boot process.
\end{itemize}
Thus, any signature produced by \ask{} would guarantee that the system
produced by the boot process includes SELinux and IMA, actively
enforcing the intended policies. Its effectiveness depends on the
appraiser having reason to believe the signing key cannot be misused
outside the context of a trustworthy system boot and our intended
software. This must rely on some trustworthy form of access control.

%


\paragraph{A limitation.} Many forms of access control rely on the use
of a secret as identification. However, when the access being
controlled is precisely access to a secret, other forms of access
control are needed, lest we start down an infinite chain of
dependencies, or compromise
Maxim~\ref{maxim:resist:transitory:corruption}. Unfortunately, with
the current versions of the Linux kernel and SELinux available to us,
it is not clear how to implement robust access control for the use of
\ask{} without using a secret. Thus, in our design described in the
next section, we are forced to compromise on
Maxim~\ref{maxim:resist:transitory:corruption}. We discuss the details
and make concrete suggestions for improvements in
Section~\ref{sec:build:up:evidence};
cf.~Section~\ref{sec:use-cases:conf:computing}.

\subsection{Layered Systems, Layered Attestations}
\label{sec:downward:layered}

Having worked our way downward from application to mechanisms
providing security and attestation, consider now the upwards
trajectory from measurements taken at startup.  Initial hardware
control starts with the {\uefi} and boot of the initial kernel and its
\texttt{initramfs}, which can be recorded in the TPM which protects
\ask{}.  These deepest components need to be the oldest measurements
in an attestation rooted in the TPM PCRs.  Clearly, no further
attestations can guarantee the {\uefi} and its firmware:  This is
where we run out of turtles~\cite{WikipediaTurtlesalltheway}.  Thus,
our adversary model must exclude supply chain attacks or reflashing
attacks against it.

For runtime remeasurement as for boot measurement, the \emph{upwards}
order is desirable~\cite{Rowe:2016wb}.  We need the kernel to be
uncompromised to believe the application measurements.  A corrupted
kernel could cause the application measurements to return the wrong
results (e.g.~by executing system calls wrong, or by corrupting
application memory).  Given the short interval, it is hard for an
adversary to corrupt the kernel after it is measured but before an
upper layer measurement.  Repairing a previously corrupted kernel to
pass a subsequent application measurement would be faster.

Hence, the dependency relation should determine a partial ordering on
the measurement sequence.  This suggests the maxim:
\begin{maxim}\label{maxim:ascend:dependencies}
  \ascenddependencies
\end{maxim}
Generally, \emph{very recent corruptions} mean corruptions during the
course of attestation, or before the sensitive operation the
attestation justifies.  We sometimes also say, corruptions in a
\emph{short period of time}.

We use the acyclic dependency maxim repeatedly in our CDS (see
e.g.~Section~\ref{sec:build:up:roots}), but here too we faced a
tradeoff given the current practical constraints
(Section~\ref{sec:build:up:lifetime},
\ref{sec:use-cases:conf:computing}).

\subsection{Inferred Adversary Model}
\label{sec:downward:adversary:model}

Against an all-powerful adversary, there is no hope of providing
trustworthy attestations.  However, our discussion allows us to define
a very strong adversary model, against which we can still provide
trustworthy attestations.

\begin{renumerate}
  \item We assume that the adversary is unable to modify the hardware
  of our platform---including the \uefi{} firmware that
  initiates boot---whether after deployment or via supply-chain
  modification.\label{clause:adv:hardware}
  \item The adversary may log in to our system, possibly even as root,
  and initiate processes including command shells.  Thus, the
  adversary may remove, insert, and modify files in many parts of the
  filesystem, including files marked as executable.
  \item The SELinux policy defines domains in which the adversary,
  even as root, cannot run programs during a boot cycle, and any parts
  of the filesystem it cannot modify.  Other SELinux resources such as
  sockets are similar.  We assume that the boot-time SELinux policy is
  set to be immutable; IMA policies can only be tightened during
  runtime.  Both of these apply only when no kernel corruption has
  occurred.
  \item The adversary may boot the system into a state of its choice.
  However, the TPM PCR state will differ from our expected values
  throughout that boot cycle.\label{clause:adv:boot-pcrs}
  \item The adversary may observe messages on the network and deliver
  messages of its choice to our system.  So the adversary may engineer
  inputs to exercise any misbehavior of which our executables are
  capable.\label{clause:adv:dolev-yao}
  \item The adversary may corrupt a running Linux kernel, and also
  restore it to a non-corrupted state.  The restoring may succeed in a
  short period of time.\label{clause:adv:fast:repair}
\end{renumerate}
A \emph{short period of time} means the length of time between the
beginning of an attestation and the completion of the sensitive
operation the relying party will perform.

We distinguish a few finer types of adversary.  If the corrupting can
also, like the repair, occur in a short period, we call that a
\emph{fast} adversary.  If an adversary is fast and able to choose
when to do a fast corruption based on our attestation schedule, then
we call it a \emph{fast attestation-triggered} adversary.
\begin{renumerate}
  \item We assume the adversary is not a fast attestation-triggered
  adversary.\label{clause:adv:not:fast:triggered}
\end{renumerate}

LKIM relies on information provided by the kernel about itself, so a
rootkit may be able to feed LKIM misinformation if it is specifically
designed with the details of LKIM in mind.  Similarly, since IMA is
part of the kernel, and its active policy is stored in kernel space, a
rootkit specially designed to undermine our use of IMA could rewrite
the policy, or turn IMA off.  Since IMA ensures that the executable
launched to run as LKIM is the intended program, this could be part of
an attack to make our measurements unreliable.  Thus, LKIM and the
kernel (with its IMA component) form a \emph{measurement cycle}
(Section~\ref{sec:downward:layered}).  In
Section~\ref{sec:downward:underlying} we also mentioned a limitation
that required us to store a secret to authorize use of a TPM key.  If
an adversary compromises the kernel temporarily, they may be able to
locate that secret and learn its value to use repeatedly
(Section~\ref{sec:downward:underlying};
cf.~Sections~\ref{sec:build:up:evidence},
\ref{sec:use-cases:conf:computing}).
An adversary that can craft unusual corruptions based on these two
limitations in our adherence to our own maxims with the current
mechanisms is \emph{{\attestationattuned}}.
\begin{renumerate}
  \item In the current architecture of our CDS example, we assume the
  adversary is not {\attestationattuned}.\label{clause:adv:not:tuned}
\end{renumerate}
However, assumption (\ref{clause:adv:not:tuned}) could be lifted if
the architecture used virtualization to separate LKIM from the kernel
in the attestation target, and if a different method is used to
authenticate commands from the attestation infrastructure to the TPM
(Sections~\ref{sec:build:up:lifetime}--\ref{sec:build:up:evidence}).

Clause~(\ref{clause:adv:not:fast:triggered}) is different:  The advice
in Maxim~\ref{maxim:ascend:dependencies} would be pointless against
fast attestation-triggered adversaries.

%
%



\section{Attestation Structure via the Maxims}
\label{sec:build:up}

We now provide some additional detail on our CDS example system's
security and attestation mechanisms, illustrating how the maxims and
adversary model of Section~\ref{sec:downward} shaped them.  There are
four layers to this architecture:
\begin{enumerate}
  \item The boot-time, root-of-trust measurements;
  \item The kernel-level security and integrity mechanisms SELinux and
  IMA with their policies;
  \item The attestation manager (AM); and 
  \item The attestation service providers (ASPs), such as LKIM, the
  application measurers, and the \texttt{tpm\_sign} service.  
\end{enumerate}
Some additional detail is given in the appendices.  

\subsection{Booting from Roots of Trust:  Linear dependencies}
\label{sec:build:up:roots}
We rely on two \emph{hardware roots of trust}, the Trusted Platform
Module (TPM), which is the \emph{root of trust for reporting and
  storage}, and the Unified Extensible Firmware Interface ({\uefi}),
which is the \emph{root of trust for measurement}.  The {\uefi} starts
the boot process and makes the first boot measurement, so it must be
trusted, as reflected in our adversary model
clause~\ref{clause:adv:hardware}.  The {\uefi} hashes the code of a
small shim executable, together with a boot security configuration,
and deposits them into two PCRs on the TPM (see
Table~\ref{tab:pcr:usage}).
\begin{table}
  \caption{PCR usage at start-up}
  \footnotesize
  \centering
  \begin{center}
    \begin{tabular}{c@{\qquad}c@{\qquad}c@{\qquad}l}
      \pcraction{Actor}{Target PCR}{Value}{} \\
      \pcraction{{\uefi}}4{$\hash{\mbox{\texttt{shim}}}$}{} \\
      \pcraction{{\uefi}}7{$\hash{\mbox{\texttt{sec\_boot\_cfg}}}$}{} \\
      \pcraction{\texttt{shim}}4{$\hash{\mbox{\grub{}}}$} \\  \\ 
      \pcraction{\grub{}}8{$\hash{\mbox{\texttt{argv}}}$} \\
      \pcraction{\grub{}}9{$\hash{\mbox{\texttt{initramfs}}}$} \\
      \pcraction{\grub{}}9{$\hash{\mbox{\texttt{kernel}}}$} \\ \\
      \pcraction{\texttt{pol\_scr}}{11}{$\hash{\mbox{\texttt{systemd}}}$}{} \\
      \pcraction{\texttt{pol\_scr}}{11}{\hash{\mbox{\texttt{SELinux\ policy}}}}
      \\  
      \pcraction{\texttt{pol\_scr}}{11}{\hash{\mbox{\texttt{IMA\ policy}}}}
    \end{tabular}
  \end{center}
  This yields the pre-release configuration-compliant PCR values.\\
  Cf.~Table~\ref{tab:post:release:pcrs} for post-release values.
  \\[1mm] \hrule
  \label{tab:pcr:usage}
\end{table}
The shim is needed because it is signed by Microsoft, as standard
{\uefi}s require; the shim's functionality is simply to hash the code
of \grub{}, the Linux boot loader, extending a PCR with this value,
before invoking \grub{}.  \grub{} in turn will hash the argument vector with
which it will invoke its successor; it hashes the contents of the
initial file system in \texttt{initramfs}; and hash the code of the
kernel it will invoke, extending PCRs with these values.  Then it
invokes an initial program in \texttt{initramfs}, which in turn
executes a policy script \texttt{pol\_scr}, which is responsible for
hashing the system daemon code, the SELinux configuration, and the IMA
configuration; PCRs are extended with these values.  Finally,
\texttt{systemd} runs and installs SELinux and IMA with their
configurations.  The configurations will be unchangeable in the kernel
(absent kernel corruption).

Each stage of this process depends on its predecessors in the sense
that earlier misbehavior would allow the wrong code or arguments to be
invoked at a later stage.  The successive PCR values tell us what code
was in control at each stage; our inspection of that code justifies
the claims we made in the previous paragraph about the behavior of
each stage.  The code during this boot phase executes without inputs
from the outside world, meaning that data-driven corruption is not yet
a threat.  Thus, this phase provides a clean simple illustration of
Maxim~\ref{maxim:ascend:dependencies}.

\subsection{Kernel Security Policy:  Creating separation}
\label{sec:build:up:security:policy}

Promptly after the kernel is started up, it begins enforcing its
SELinux and IMA policies.  IMA can ensure that critical executables
will be launched as processes only if their computed hash matches the
intended value.
%
%
Our IMA policy covers all of the executables that make up our
application, from \resizesf{intake} through \resizesf{rewrite} and
\resizesf{filter} to \resizesf{export}.  They also include the paths
to the Attestation Manager and various Attestation Service Providers.
These values are checked before \texttt{exec}ing the files at these
paths.

The SELinux policy associates security contexts with executing
processes as well as with system resources such as files, directories,
sockets, and so on, and constrains interactions depending on the
source and target security context.  Our SELinux policy protects:
\begin{itemize}
  \item the integrity of the attestation mechanism, and
  \item the integrity of the CDS application and the confidentiality
  of the messages it handles.
\end{itemize}
Thus, it is the key mechanism for
Maxim~\ref{maxim:constrain:interaction}.  To justify the backbone of
trust, the appraiser needs only a collection of evidence showing that
(i) our SELinux policy is active, (ii) the processes with AM or CDS
contexts are running the intended executables, and (iii) the
measurements report acceptable values.  Since we will always set the
SELinux policy to be immutable, we know that boot-time SELinux policy
remains in force during runtime, in the absence of kernel intrusions.
We remeasure it when attesting, as well as the IMA policy.  

\paragraph{Policy for Attestation Manager.}  The policy for
attestation must ensure its integrity.  It stipulates security
contexts for the executable files and for the processes that execute
them.  No system activity other than a system management role should
be able to modify them.  The Attestation Service Providers (ASPs) can
be invoked only by the Attestation Manager (AM); the results can be
passed back only to the manager; and the manager passes the partial
evidence only to subsequent services and completed evidence only to a
socket for delivery to the appraiser.

\paragraph{Policy for the CDS}  The CDS policy enforces the pipeline
flow (Fig.~\ref{fig:cds}).  Filesystem paths are labeled to ensure
each pipeline stage can \texttt{exec} only the file at the intended
location; IMA will ensure this is the correct executable.  No system
activity other than a system management role should be able to modify
them.  Each pipeline process can read only from its intended source,
and write only to its intended pipeline target, together with an
additional directory for logging and errors.  This condition also
ensures that unexpected behavior in a pipeline process cannot lead to
confidentiality failures that leak messages to other activities on the
CDS platform.

Thus, the SELinux and IMA policies provide strong kernel-level
enforcement for the isolation that
Maxim~\ref{maxim:constrain:interaction} requires.  

\subsection{Short-lived vs.~Long-lived}
\label{sec:build:up:lifetime}

We have designed the CDS application processes to be short-lived, in
accordance with Maxim~\ref{maxim:short:lived}, since their
functionality is compatible with a single-message lifetime.  The
server process listening on the intake socket forks a new process for
each message; that process writes it to a file in the incoming
directory and then invokes a rewrite process with that filename as
argument, and so on.  As a consequence, there is no need to worry that
a maliciously constructed message could cause subsequent messages to
be rewritten or filtered incorrectly.

The attestation mechanism is constructed similarly.  A server
listening on a socket forks a new AM process for each incoming
attestation request.  That server forks and \texttt{exec}s ASPs as
needed to perform measurements and also the signature requests for the
TPM to apply the Attestation Signing Key $\ask{}$ to each bundle of
evidence as it becomes ready.  Each of these processes executes one
measurement or obtains one signature, returning its result to the AM
and exiting.  Since the IMA policy ensures that the intended
executable is at each of the intended paths, and the SELinux policy
controls which processes can invoke them (or be invoked), the correct
execution behavior is assured.

By contrast, the Linux kernel is long-lived, and could not be replaced
by a short-lived execution monitor in practice.  Thus, by
Maxim~\ref{maxim:short:lived}, a remeasurement technique is indicated.
LKIM, the Linux Kernel Integrity Measurement tool, meets this need.
It systematically re-evaluates data structures (including the system
call table and other function pointers) and loaded code to ensure that
the invariants on which normal kernel behavior depends are satisfied
at the time of remeasurement.  In our demonstration case study, this
is the only address space that we inspect to revalidate invariants.

Except in these special cases, relying on short-lived processes is the
practical alternative.  Hence, the long-lived branch of
Maxim~\ref{maxim:short:lived} will rarely be preferred above the
kernel level.

\paragraph{A tough choice.}  The current version of LKIM uses eBPF,
the extended Berkeley Packet Filter, to obtain information from the
running kernel from which it reconstructs the invariants.

Older versions used a different strategy.  If a kernel was running as
a virtual machine atop the Xen hypervisor, LKIM ran as a separate
virtual machine with read-only access to the target for virtual
machine inspection~\cite{thober2008improving},~\cite[\S
5.1]{Coker::Principles-of-R}.  This had an advantage:  Even a
sophisticated rootkit in the kernel under measurement cannot falsify
the data LKIM received.  However, Linux is rarely run as a virtual
machine on Xen currently.  Various virtualization mechanisms are
currently available---making it hard to select a single approach---and
the Linux kernel is often run unvirtualized.

Misleading LKIM would require a very sophisticated rootkit,
identifying which eBPF queries should receive false responses.  Thus,
although this strategy diverges from
Maxim~\ref{maxim:ascend:dependencies}, since the measurer depends on
the target of measurement, we consider this practically realistic and
sufficiently difficult to mislead, absent a sophisticated
attestation-attuned adversary
(Section~\ref{sec:downward:adversary:model}
clause~(\ref{clause:adv:not:tuned})).
\begin{recomm}
  \label{recomm:lkim:acyclic}
  A new version of LKIM should be able to run as a separate virtual
  machine on a widely available hypervisor, receiving kernel data via
  VM inspection.
\end{recomm}

\subsection{Providing Evidence}
\label{sec:build:up:evidence}

When the Attestation Manager has assembled its evidence, it needs to
use the $\ask{}$ to apply a digital signature so that the appraiser
can infer it was correctly generated
(Maxim~\ref{maxim:display:boot:trait}).  Signing must be unavailable
unless the platform is currently executing under a boot complying with
our intent, and in particular with our SELinux and IMA policies.
Maxim~\ref{maxim:resist:transitory:corruption} also stipulates that
the $\ask{}$ should not be resident in the AM's address space, since
then a transient corruption would expose the $\ask{}$ and allow the
adversary to sign at will thereafter.  

To meet this requirement, we arrange for the $\ask{}$ to be a TPM
signing key.  It is usable only within the TPM, and outside the TPM it
is always wrapped (i.e.~encrypted).  It is subject to a TPM key policy
that stipulates it can be used only when the TPM PCRs 4, 7, 8, 9, and
11 contain the values resulting from the boot process of
Table~\ref{tab:pcr:usage} applied to our selected values of the
parameters shown there.  Updating our intended configuration on the
platform requires revoking the old $\ask{}$ or allowing its
certificate to expire, and then provisioning a new key with the new
key policy.

\paragraph{A fly in the ointment.}  Crucially, the TPM must not use
the $\ask{}$ to sign evidence unless the AM (or its \texttt{tpm\_sign}
ASP, in fact) makes the request.  A rogue process on the platform
should never succeed when submitting a request for a value to be
signed with this key.  How can this be ensured?

Here there are two strategies.

\paragraph{Using TPM localities.}  The natural strategy would be to
use the TPM localities.  TPM localities are restrictions on the source
of commands.  Some are associated with specific hardware sources, such
as early boot events, but others are flexible.  The $\ask{}$ key
policy could embed a locality restriction.  This could be combined
with an SELinux rule requiring that the kernel deliver a TPM command
with that locality only if the source is a process running with a
particular security context.  In our case, that would be the security
context of the \texttt{tpm\_sign} ASP, which with IMA's help we
already know is running only with evidence the AM intends to receive
an $\ask{}$ signature.  In more detail:
\begin{enumerate}
  \item \ask{} is wrapped with a key policy stipulating that each
  signing command require a configuration-compliant PCR state $P_A$,
  and that the command should be delivered to the TPM labeled with a
  particular locality $A$.
  \item The kernel, when running our SELinux policy, labels commands
  with locality $A$ only when the request is delivered by a process
  running in domain $D_A$.
  \item Our SELinux policy assigns domain $D_A$ only to the executable
  at a particular path in the filesystem, \texttt{tpm\_sign} in our
  case, which can be invoked only from a domain $D_{AM}$ that only AM
  instances run within.
  \item Our IMA policy enforces that the \texttt{tpm\_sign} and AM
  executables have the hashes determined for the correct signing
  program and Attestation Manager.
  \item The PCR state $P_A$ incorporates our SELinux and IMA policies.  
\end{enumerate}

Unfortunately, this ``natural approach'' is hypothetical.  The Linux
kernel has no built-in API to deliver commands with TPM \emph{extended
  localities}, i.e.~localities $\ell$ with $32\le\ell<256$, although
an alternative TPM resource manager such as IBM's
one~\cite{GoldmanIBMswTPM2019} can give some control over extended
localities.  These are currently reserved by the Trusted Computing
Group~\cite[\S 3]{TCGReservedLocalities2019}, but some could better be
used to indicate the security-relevant identity of the source process
of a command.  This functionality would also need Linux Security
Module hooks to allow policies to control the source of commands to
the TPM.  SELinux would also need to include TPM localities as an
object type that it can protect via its rules.  This enrichment to
Linux Security Modules and SELinux would be a natural way to make the
TPM a more robust security mechanism for many purposes beyond our
attestation goals.  We recommend that the kernel and SELinux be
extended to do so.
\begin{recomm}
  \label{recomm:tpm:localities}
  The Linux kernel should have a standard way to deliver commands from
  user level processes to the TPM with specific extended localities
  $32\le\ell<256$.  SELinux should support this by implementing TPM
  localities as an object class, allowing configurations to determine
  which process security contexts may deliver commands with any given
  software locality.
\end{recomm}
We used an alternative strategy, not having a realistic way to modify
the kernel and SELinux.  It is less direct, and it diverges from
Maxim~\ref{maxim:resist:transitory:corruption}, the maxim on avoiding
memory-resident secrets whose disclosure would undermine attestation.
This stopgap strategy does not use TPM localities.

\paragraph{Key policies as secrets.}  Our alternative uses a secret
that is either a key blob or the policy for using that key blob.  

TPM key usage policies, which stipulate PCR values and localities that
must be active for the key to be used, come in two forms.  An
\emph{immediate} key usage policy is packaged with a TPM key such as
\ask{} and always accompanies it, even when the key is wrapped and
transmitted externally as a key blob.  Alternatively, a key may have a
\emph{delegated} policy.  In this case, a signature verification key
or \emph{delegation key} accompanies the TPM key when resident or
wrapped in a blob.  A \emph{delegated key usage policy} is a signed
data structure containing a usage policy.  If it verifies under the
delegation key, then the TPM will use the key under that policy.
Delegated policies are useful when developing a system, or if the
desired PCR values will change relatively frequently, since otherwise
the system manager would have to reprovision the system with new TPM
keys with a new immediate policy.

We first describe an alternative using TPM keys with immediate
policies.  We maintain the wrapped \ask{} with its immediate policy
doubly wrapped with a second TPM-enforced encryption layer (via a LUKS
container).  The TPM policy for this secondary encryption is subject
to a key policy that stipulates that it can be decrypted only when the
PCRs are in the ``pre-release configuration-compliant'' state defined
in Table~\ref{tab:pcr:usage}, meaning that IMA and SELinux are running
with our policies in force.  The policy script \texttt{pol\_scr} has
the TPM unwrap the outer layer; \texttt{pol\_scr} places the singly
wrapped \ask{} key blob into the filesystem with a security context
ensuring that only \texttt{tpm\_sign} can read it.  Then it updates
the PCR 11 state (Table~\ref{tab:post:release:pcrs}).

\texttt{tpm\_sign} delivers this key blob in a TPM authenticated
session along with each signing request.  The TPM does not cache it.
Thus, access to the key blob authenticates the requester as
\texttt{tpm\_sign}.

If using delegated key usage policies, we need not double wrap the
actual key blob with its delegation key.  Instead, we wrap the
delegated key usage policy using a TPM key that can be used in the
pre-release configuration-compliant state.  This delegated key usage
policy is placed in the filesystem for the exclusive use of
\texttt{tpm\_sign}.  The delegated key usage policies ease system
management, but come with a security risk:  One must never use the
same signing key to certify policies for two different keys that may
have different purposes.  Otherwise, the policy meant for one key
could be misused with the other key.   

Either version of this mechanism does roughly comply with
Maxim~\ref{maxim:display:boot:trait} since the key blob will never
become available in an untrustworthy boot cycle.  If nothing else has
gone wrong, in a trustworthy boot cycle it is readable only by
\texttt{tpm\_sign}.  However, it does not follow
Maxim~\ref{maxim:resist:transitory:corruption}:  A transient
corruption of the kernel would allow an adversary to read the
\texttt{tpm\_sign} process memory that holds the key blob.  It would
also allow the adversary to bypass the SELinux rules and directly read
the file containing the key blob.  Having obtained this value, the
adversary could patch the kernel to allow future LKIM runs to report
an uncorrupted state.  Nevertheless, the adversary could use the key
blob at will when running other user-level processes.
\begin{table}
  \caption{After releasing the \ask{}}\vspace{1mm}
  \footnotesize
  %
  \centering
  \begin{center}
    \begin{tabular}{c@{\qquad}c@{\qquad\qquad}c@{\quad\qquad}c}
      \pcraction{Actor}{Checking}{Target PCR}{Value} \\
      \pcraction{\texttt{pol\_scr}}{4, 7, 8, 9, 11}{11}{\texttt{"ok"}}
    \end{tabular}
  \end{center}
  This yields the post-release configuration-compliant PCR values.\\[1mm] \hrule
  \label{tab:post:release:pcrs}
\end{table}
Hence, we regard this as a stopgap, and we believe that
Maxim~\ref{maxim:resist:transitory:corruption} provides a strong
reason for implementing TPM localities in the Linux kernel and
SELinux.



\section{Implementation and Evaluation}  
\label{sec:self:test}

How trustworthy is our attestation design?  There are two
different ways it could be untrustworthy.  For one, the
measurements we take could be unreliable.  An adversary could
cause a particular component to misbehave, and if our measurement
does not reveal that, then we should not believe the attestation it
is part of.

Second, there could be structural problems in the way our
attestation is put together.  For instance, suppose we measured
the \resizesf{rewrite} configuration by hashing the file at the
expected path, followed by a sequence of other measurements before
finally using LKIM to check the kernel.  An adversary who had
corrupted the kernel could replace the good file at that location
after the hash with a bad configuration followed by a fast kernel
repair.  This is compatible with the adversary model of
Section~\ref{sec:downward:adversary:model},
item~(\ref{clause:adv:fast:repair}).  However, if the LKIM check
happens before, then the kernel corruption would have to take
place between this check and the configuration file measurement.
This fast corruption would be challenging in reality, and
item~(\ref{clause:adv:not:fast:triggered}) records our assumption
that corruption takes craftsmanship and therefore time.

We start in Section~\ref{sec:self:test:impl} with background
on how we implement attestation.
In Section~\ref{sec:self:test:layers} we consider the first type
of failure, using testing to ensure that our individual
measurements give good results.  In
Section~\ref{sec:self:test:formal} we consider the second, using a
formal analysis to explore the attestation structure.

\subsection{Prototype Implementation}
\label{sec:self:test:impl}

The collection of software components that facilitate the transition
from boot-time to run-time attestation is the \emph{Attestation
  Manager Subsystem}.  Among these is a server process called the
\emph{Attestation Manager Launcher} that listens for (run-time)
requests and launches an \emph{Attestation Manager} (AM) instance, which
in turn orchestrates auxiliary executables called \emph{Attestation
  Service Providers} or ASPs to gather and bundle runtime evidence.
We instantiate the Attestation Manager using the
  {\maestro}\footnote{Measurement and Attestation Execution and
  Synthesis Toolkit for Remote Orchestration.} toolkit for layered
attestation
systems~\cite{Petz:2024:Verified-Configuration-and-Deployment-Paper}.
{\maestro} components on the CDS target platform orchestrate access to
measurement primitives like the LKIM back-end and the TPM signing
capability.
  %
  %
  %
  %
  {\maestro} is also able to generate \emph{appraisal} components to run
on an appraiser.  These appraisal components unbundle the evidence
that the \emph{attestation} mechanism built, and evaluates their
structure and the individual measurements they contain.  These
corresponding attestation and appraisal procedures ensure that all the
evidence gathered is interpreted correctly by the appraiser to yield
the right trust decision.  {\maestro}-synthesized components adhere to
the formal semantics of the Copland~\cite{Ramsdell:2019aa} domain
specific language for attestation protocols, with accompanying proofs
of correctness in the Rocq proof assistant.
%

\paragraph{Performance Benchmarking}

Our prototype implementation is deployed on a Fedora 41 Linux desktop running an Intel Xeon E-2124 4-core CPU with 16GB RAM and a Samsung PM981 NVMe SSD.  For ease of development we use the \texttt{swtpm} TPM emulator \cite{swtpm}.  We also use the standard installations of IMA and SELinux packaged with Fedora 41.  The only customizations to this commodity Fedora install are the hook scripts added to \textsf{initramfs} to measure IMA and SELinux policy configurations to TPM PCRs during boot.  The CDS pipeline components are implemented in the memory-safe language OCaml.

\ifrotatedtables
\begin{landscape}
\fi
\begin{table*}[htbp]
  \centering
  \caption{Performance Benchmarks: Baseline vs. AM Running (varying $\Delta$)}
  \label{tab:benchmark-results}\small
  \begin{tabular}{lrrrrr}
    \toprule
    \textbf{Metric}                             & \textbf{Baseline} & \textbf{AM ($\Delta=60$s)} & \textbf{AM ($\Delta=30$s)} & \textbf{AM ($\Delta=15$s)} & \textbf{\% Reduction} \\
    \midrule
    \textbf{CDS Throughput (over 600s period)}  &                   &                            &                            &                            &                       \\
    \hspace{1em} Total Requests Processed       & 84060             & 83328                      & 83040                      & 82956                      & 1.3\%                 \\

    \addlinespace
    \textbf{OSBench (Lower Better)}             &                   &                            &                            &                            &                       \\
    \hspace{1em} Create Threads ($\mu$s)        & 19.18             & 18.96                      & 18.81                      & 19.03                      & -0.8\%                \\
    \hspace{1em} Launch Programs ($\mu$s)       & 120.25            & 122.43                     & 121.49                     & 122.01                     & 1.4\%                 \\
    \hspace{1em} Create Processes ($\mu$s)      & 32.85             & 33.38                      & 33.47                      & 34.31                      & 4.3\%                 \\
    \hspace{1em} Memory Allocations (ns)        & 87.62             & 88.21                      & 88.62                      & 87.78                      & 0.2\%                 \\

    \addlinespace
    \textbf{System Compilation (Lower Better)}  &                   &                            &                            &                            &                       \\
    \hspace{1em} Linux Kernel 6.15 (s)          & 775.51            & 777.73                     & 786.70                     & 793.59                     & 2.3\%                 \\

    \addlinespace
    \textbf{Server Performance (Higher Better)} &                   &                            &                            &                            &                       \\
    \hspace{1em} nginx 1.23.2 (req/s)           & 9845.89           & 9828.28                    & 9774.56                    & 9651.66                    & 0.7\%                 \\
    \bottomrule
  \end{tabular}
\end{table*}
\ifrotatedtables
\end{landscape}
\fi

We benchmark our system utilizing Phoronix Test Suite
\cite{phoronix_test_suite}, a widely used benchmarking framework for
Linux systems.  We compare the performance of our system with and
without the AM running, varying the attestation
interval $\Delta$ (i.e., how often runtime attestation is performed)
among 60s, 30s, and 15s.  The benchmarks we run include: general OS
tasks, building the Linux kernel from source, Nginx web server
performance, and overall CDS processing throughput over a 10-minute
period.  The results are summarized in
Table~\ref{tab:benchmark-results}. The results indicate that the
performance overhead introduced by the Attestation Manager is minimal
across all benchmarks, even at the most frequent attestation interval
of 15 seconds. In particular, the throughput of the security critical
application (the CDS pipeline) remains high, with only a 
  %
  %
  {1.3\% reduction in processed requests} when the AM is running with a
15-second attestation interval compared to the baseline without
attestation.

We further experimentally establish an average time it takes for the
attestation of the CDS system complete: 
{5.48 seconds}. Of this, 
{5.05 seconds} (or 92.2\%) was dedicated to the LKIM measurement
procedure. This was established by running the attestation procedure
100 times and averaging the total time taken.

\subsection{Empirical Testing}
\label{sec:self:test:layers}



In order for a relying party to trust the appraisal of the CDS system, they must also trust the \emph{measurers} it uses to detect and report corruptions.
For example, maintaining the integrity of a CDS executable in User Space requires trust that IMA will reliably measure, detect, and prohibit malicious updates to that file.
In turn, trusting IMA requires being able to detect corruptions in the Linux kernel and its boot process.
In what follows, we provide empirical justification that the measurers we use can detect typical attacks against target components.
Further, we comment on why we expect other attacks to also be detected, when the components the measurers depend on are uncorrupted.

\newcommand{\yes}{\textcolor{green}{\ding{52}}} 
\newcommand{\no}{\textcolor{red}{\ding{56}}}   
\newcommand{\na}{\textcolor{gray}{N.A.}}   

\ifrotatedtables
  \begin{landscape}
  \fi
\begin{table*}[tbp]\small
  \begin{tabular}{l l l}
    \toprule
    \textbf{Components}
    & \textbf{Attacks}
    & \textbf{Defense} \\
    \midrule

    \textbf{Hardware (CPU, TPM, etc.)}
    & \emph{Outside of Adv. Mod.}
    ~\S\ref{sec:downward:adversary:model}, item~(\ref{clause:adv:hardware})
    & ---

    \\ \addlinespace

    \textbf{Boot Process}
    & Corrupt boot scripts, IMA/SELinux policies
    & Signing key unavailability; TPM quote
    \\ \addlinespace

    \textbf{LKIM}
    & Corrupt LKIM binary
    & IMA policy (prevents launch)
    \\ \addlinespace
    & Corrupt kernel--LKIM cycle
      & ---$^\dagger$\\
    & \quad \emph{Outside of Adv. Mod.}~\S\ref{sec:downward:adversary:model}, item~(\ref{clause:adv:not:tuned})
    \\ \addlinespace

    \textbf{Kernel (incl. IMA/SELinux)}
    & Corrupt kernel, modules
    & LKIM appraisal
    \\ \addlinespace

    & Corrupt IMA or SELinux policies
    & Immutable (up to kernel corruption)
    \\ \addlinespace

    \textbf{User Space (AM, ASPs, CDS)}
    & Corrupt executables, configs
    & IMA/SELinux policies; 
     application-level measurers
    \\ \addlinespace

    & Send malicious inputs
    & Short-lived processes \\

    \bottomrule
  \end{tabular}

  \caption{Components, attacks on them, and defense mechanisms for our CDS system.\\
  $^\dagger$In a virtualized setting (as described in Section~\ref{sec:use-cases:conf:computing}), this cyclic limitation can be lifted by isolating LKIM via a hypervisor.}
  \label{tab:attacks}
\end{table*}

\ifrotatedtables
\end{landscape}
\fi



In Table~\ref{tab:attacks}, we summarize attacks against different layers of the CDS target system and mechanisms by which they are detected by our layered attestation system.
The first column lists the major component classes, ascending from
the deep Hardware mechanisms to the shallow User Space components.
The other two columns indicate
%
%
kinds of
attacks and corresponding
%
%
defenses
for each component class.

\paragraph{Boot-time Attacks}

The 
%
%
deepest
level of integrity violations detected by our attestation framework
occur during system boot.  Attacks that target firmware are out of
scope as {\uefi} is considered a root of trust according to item
(\ref{clause:adv:hardware}) of the inferred adversary model.  The Boot
Process component in Table~\ref{tab:attacks} is shorthand for the
configurable components involved in boot, including the \grub{}
bootloader, custom boot scripts, and their data dependencies.  In
testing, we used the measurement targets in the ``Value'' column of
Table~\ref{tab:pcr:usage} to craft our example attacks.  We first
performed a ``provisioned boot''---into an assumed secure state---to
record golden values for each target.  In subsequent boots, we
modified each target in turn and validated that the key release scheme
of Section~\ref{sec:build:up:evidence} denied
access to the ASK signing key because the PCR values were different.

\paragraph{Kernel-level Attacks}

As a representative example of an attack against the Linux kernel, we
used an out-of-the-box kernel module insertion attack
\cite{invary_test_kit}.
%
%
Once it is loaded, the next LKIM measurement disclosed the
corrupted kernel.
As another example of a kernel-level attack, we replaced the installed
CDS SELinux policy with a modified one while the kernel was corrupted.
After
%
%
this modification, the SELinux policy measurement ASP disclosed the
alteration.
%
%
If, however, a kernel corruption disables IMA, which we demonstrated
can occur, then the LKIM executable or SELinux policy measurement ASP
could be replaced with a dummy that would always return acceptable
measurements.  This is an example of an attack by an
  {\attestationattuned} adversary, which we exclude from our adversary
model in Section~\ref{sec:downward:adversary:model},
item~(\ref{clause:adv:not:tuned}).


\paragraph{User-level Attacks}

%
%
%
The CDS components, namely the pipeline executables and their
configuration files, are prime targets for the adversary at the
user-level layer of the system.  The adversary wants to replace
legitimate CDS components with buggy or malicious versions while
keeping the CDS operational, but avoiding detection by the attestation
subsystem.  Our layered attestation verifies the identity of these
components using a binary hash equality.

%


We executed attacks that swap out each of the CDS user-level
components in turn with a malicious version that violates CDS security
goals.
%
%
Our attestation protocol detected unrepaired changes to these
components.  Our IMA policy, with its integrity checks at process
launch time, provides additional protection by preventing such
modified programs from running altogether.

\paragraph{Attacks on the attestation mechanism.}

We explored various attacks on the attestation mechanisms themselves.
First, we attempted to replace or modify the core AM and ASP
executables that carry out the CDS attestation protocol. These attacks
are prevented by IMA, just as the CDS user-level components are.

Next, by adversary model item~\ref{clause:adv:dolev-yao}, an adversary
can deliver evidence to the appraiser that was not generated recently
in the system's attestation machinery, or was not generated there at
all.

An attacker can also store and replay the signed evidence package from
a previous run of attestation on the target.  However, in our
attestation protocol the appraiser delivers the query with a freshly
chosen nonce, which is embedded in the signed evidence package.  Thus,
the appraiser will reject a replayed evidence package, which contains
a nonce that does not match its current expectation.

An adversary operating outside the system---or its attestation
machinery---faces the task of signing evidence with the $\ask{}$,
without which the appraiser will not accept it.  By our threat model,
this key is never usable except within the system's TPM.  Moreover, it
can be used only if accompanied by the wrapped key policy
(Section~\ref{sec:build:up:evidence}), which is available (in the
absence of a kernel corruption) only to the TPM signing ASP.  Thus,
within the scope of our adversary model,
item~(\ref{clause:adv:not:tuned}), the adversary will not corrupt the
kernel and also successfully retrieve the wrapped key policy.  Thus,
the forged evidence will not be signed by the $\ask{}$, and the
appraiser will reject it.  
We did not attempt to execute these two attacks.

\subsection{Formal Analysis}
\label{sec:sefl:test:formal}
\label{sec:self:test:formal}

The formal analysis uses the techniques and tools of Rowe
et~al.~\cite{Rowe:2016wb,Rowe-Automated-Trust-21}.  Our attestation
protocol is written in the Copland language, whose formal semantics
determines a partial order on the events such as measurements that the
protocol requires.  The analysis for a particular query systematically
explores all ways the adversary could interpose corruption and repair
events in this partial order subject to rules that reflect the
adversary model and the structure of the system.  It reports cases
where the component mentioned in the query would be reported as
acceptable when it was in fact corrupted.  The analysis reflects
dependencies, for instance the dependency of a user-level process on
the kernel, because a user-level measurement process may report that
its target is in an acceptable state when it is in fact corrupted, if
the underlying kernel is corrupted when it runs.

We performed several analyses by running the tools on a model of our
system design under varying sets of assumptions. This section
summarizes the inputs and the results of those analyses at a high
level.

Our system design involves integrity measurements of various
components during both boot and runtime. Each of these measurements
consists of a \emph{measurer} measuring a \emph{target}. We assume
that if the target is compromised but the measurer is not, then
measurement will result in evidence that the appraiser will not
accept. In other words, an uncompromised measurer is capable of
detecting a compromised target. On the other hand, if the measurer is
compromised by the adversary, then the adversary has the incentive and
the ability to fabricate evidence that will pass
appraisal. Additionally, a measurer may depend on one or more other
components to perform a reliable measurement. For example, if the
kernel is compromised, then user space ASPs that rely on it will not
function as expected, even though their executable code may be
unaltered. Thus, even when a measurer is uncompromised, if it depends
on a compromised component, we assume the adversary has the incentive
and ability to use the compromised component to produce evidence that
will pass appraisal.

Given a collection of measurement events temporally ordered in some
specified way, a relying party of an attestation wants to know: Is it
possible for some component---for example, the CDS code---to be
compromised when it is measured without the appraiser receiving
evidence of this (or any other) compromise? The analysis proceeds by
enumerating all collections of adversary actions (subject to an
adversary model described below) that would be necessary for the
answer to that question to be ``yes''~\cite{Rowe-Automated-Trust-21,
  Ramsdell:2020:Chase}. If the analysis completes without finding any
way for the adversary to avoid detection, this is an indication that
our design is resistant to the identified adversary model. 

\begin{figure}
  \centering
  \includegraphics[scale=0.6]{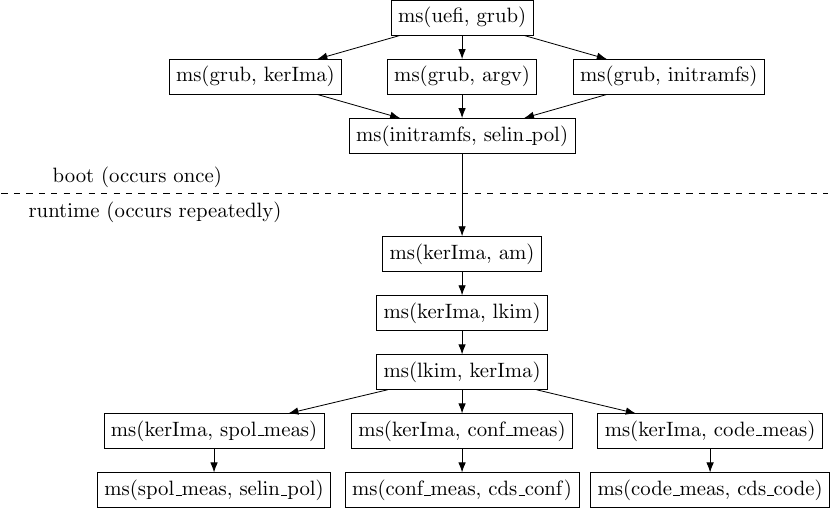}
  \caption{Temporal order of all measurements including both boot and
    runtime. Notation ms(x, y) indicates x measures y. An arrow from
    node X to node Y indicates X temporally precedes Y.}
  \label{fig:measurement:order}
\end{figure}

The attestation structure described in Section~\ref{sec:build:up}
specifies measurement events temporally ordered as shown in
Fig.~\ref{fig:measurement:order}. In the figure we write kerIma to
denote the kernel with the IMA subsystem (code and configuration) as a
subcomponent. This is because the LKIM measurement of the kernel
should also measure the IMA module and its policy. That is, corrupting
the IMA module or its policy is a way of corrupting the kernel. 

In the analysis we assume that the AM depends directly on the kernel
(kerIma) to function properly. The IMA code depends on the IMA policy
to function correctly. This dependency is naturally encoded by viewing
IMA and its policy as being included in kerIma.
Finally, we assume that each of the ASPs depends on the kernel because
a corrupt kernel could alter memory in such a way to cause an ASP to
measure a target of the adversary's choosing.

The default adversary model used in prior
work~\cite{Rowe-Automated-Trust-21} assumes the adversary can
compromise any component (including the kernel) at any time.  This is
consistent with items (2), (4), and (6) of the adversary model in
Section~\ref{sec:downward:adversary:model} which assumes the adversary
may gain root privileges, boot the system into a compromised state,
and corrupt the kernel. Since compromised components might be detected
by measurements, the adversary also sometimes has the incentive to
\emph{repair} a component returning it to an uncompromised
state. These repair actions can also occur at any time as per item (6)
of the adversary model. This default adversary model is stronger than
the one in Section~\ref{sec:downward:adversary:model}. As a result, we
should expect an analysis using this adversary model to return
possible attacks that defeat the attestation, and indeed it does.

The method of analysis allows us to specify patterns of attack to
exclude from the search. By doing so, we can further constrain the
adversary model to accurately align with
Section~\ref{sec:downward:adversary:model}. Item (1) states that the
adversary is unable to modify uefi, the root of trust for
measurement. We therefore exclude from the analysis any results that
contain such a corruption. Similarly, since the IMA and SELinux
policies are immutable at runtime unless the kernel is rootkitted
(item (3) of the adversary model), we impose a constraint that any
corruption of the IMA or SELinux policies must be preceded by a
corruption of kerIma. In fact, our choice to model the IMA policy as
part of the kernel already imposes this constraint for that policy. To
account for item (7) which assumes it's not a fast
attestation-triggered adversary, we exclude results containing a
corruption of some component during boot or runtime. That is, we only
consider what can happen if the adversary can corrupt components
before boot (e.g., a persistent corruption during a previous boot
cycle) or between boot and runtime (essentially at the dotted line in
Fig.~\ref{fig:measurement:order}). Each of these additional
restrictions excludes some attacks found by the original analysis.

The added constraints above do not preclude an {\attestationattuned}
adversary. As a result, we would like any remaining attack to be one
that exploits the acknowledged cycles between LKIM and the kernel
(including the embedded IMA module and policy). The cycles are of two
types. First, LKIM depends on the kernel to provide it information. So
an uncorrupted LKIM could be lied to by a corrupted kernel. If we
remove the assumed dependency of LKIM on the kernel, this will
excluded attacks that exploit that dependency. Second, there is a
measurement cycle in which the kernel (via IMA and its policy)
measures LKIM and LKIM measures the kernel (including IMA and its
policy). Exploiting this measurement cycle requires the adversary to
corrupt \emph{both} the kernel and LKIM between boot and runtime. By
finally adding a constraint that at most one of them can be corrupted
in that window, we effectively exclude an {\attestationattuned}
adversary. The resulting analysis reveals that no other attacks are
possible against the system.

As a whole, the analyses described above demonstrate that our system
design is able to withstand an adversary with the capabilities
described in Section~\ref{sec:downward:adversary:model}. They also
highlight the role played by each of the restrictions imposed on the
adversary (items (1), (3), (7), and (8) of the adversary
model). Loosening any of those restrictions results in an adversary
strong enough to defeat our attestation. While we believe the first
three are quite reasonable restrictions, item (8) is only needed due
to limitations in the available technologies that forced us to
compromise on Maxim~\ref{maxim:ascend:dependencies}. In
Section~\ref{sec:amd_use-case}, we explore a possible alternative
design that could potentially also resist an {\attestationattuned}
adversary. 



\section{Breadth of Applicability}
\label{sec:use-cases}

We formulated our maxims while working on the CDS example, in response
to the strategic choices we faced.  But in writing them, we focused on
aspects that would determine attestation trustworthiness across a
range of cases.  In this section, we will discuss two changes in the
system context, showing that our maxims continue to shape our
decisions effectively.

We have not implemented these alternate contexts, unlike the fully
implemented CDS, so our discussion will be sketchier.

\subsection{Distributed Document Control}
\label{sec:use-cases:doc:control}

Our CDS example is a single platform.  Would we need essentially
different strategies for attestation in a distributed application?
For instance, consider a system for allowing controlled access to a
collection of sensitive (possibly proprietary) documents.  The
documents normally reside on one or more platforms that serve them and
function as the document store.

They may be checked out by a collection of workstations distributed
throughout the organization's network.  While a document is checked
out, the user of that workstation may read it or may make alterations
to be checked back later.  The sensitive documents must be under the
care of specific, trustworthy application software while checked out.
A specific workflow is needed on the workstations, assuring that each
document is safely deleted when the user signals the end of the
session using it.  If the document is updated, there should be checks
to ensure its contents are well labeled before committing it back to
the document store.

The document store must ensure that it delivers documents only to
authenticated and attested workstations for authenticated end users to
use through the intended application.  The trustworthy application on
the workstation must also authenticate the document store and ensure
it is trustworthy when retrieving documents (to ensure their
integrity) and when delivering updates (to preserve their
confidentiality).  Mutual TLS is a mechanism to provide the reciprocal
authentication we need, with suitably prepared certificates.

Maxim~\ref{maxim:constrain:interaction} suggests that each of the
platforms---including the workstations and the document stores---has a
separation between the software implementing the document system and
the remainder of the platform.  We would expect to use SELinux
and IMA policies to ensure this in a way similar to the CDS.

In addition, in this context, there is an additional type of
separation that matters, namely separating the appraisal from the
primary platforms.  In order to keep appraisal separated from all of
the functionality of the system, we would run the appraiser as its own
independent platform.  The appraiser platform issues a periodic
attestation request to each active platform, namely the workstations
and document stores.

On successful completion, the appraiser, doubling as the Certifying
Authority for the mTLS certificates, issues a short-term digital
certificate for a new signing key pair generated during the
attestation process.  The signing key in this key pair will be
protected via SELinux for the exclusive use of the network component
of the application on the workstation.  This is compatible with
Maxim~\ref{maxim:resist:transitory:corruption}:  If the signing key is
disclosed in a transitory corruption, it can be misused only during
the lifespan of the short-term certificate.  After that, continued
participation in the distributed application requires a new certified
key, which will be based on attestation that uses a TPM-resident key.
This TPM-resident key will be handled in accordance with
Maxim~\ref{maxim:display:boot:trait}, i.e.~it will be available only
during a compliant boot, much as we described in
Section~\ref{sec:build:up}.  Our recommendation
(Section~\ref{sec:build:up:evidence}) that TPM localities be properly
controlled via SELinux is still needed here.

Thus, the platforms need only interact via mTLS using the short-term
certificates issued by the appraiser/CA.  This ensures that the peer
has recently passed an appraisal, at least assuming the systems have
trustworthy access to the time.  Time would have to be non-decreasing
and loosely synchronized with the appraisal/CA platform.

Maxims~\ref{maxim:short:lived} and~\ref{maxim:ascend:dependencies}
play much the same role as in our main example.

And who will appraise the appraiser?  For this, manual scrutiny may
suffice.  For one thing, the appraiser, being also a CA, should really
be used for no other purposes, this being a platform-wide application
of Maxim~\ref{maxim:constrain:interaction}.  Its boot process can be
constructed along the same lines as our main example, with the CA
signing key made available only on a compliant boot.  If the appraiser
is rebooted regularly---a variant version of the
Maxim~\ref{maxim:short:lived} idea of short-lived computational
processes---then an administrator can review its configuration on each
reboot using a script to check the hashes of executables.

\subsection{Confidential Computing as Hardware Root}
\label{sec:use-cases:conf:computing}
\label{sec:amd_use-case}

In our main example, we used a widely available hardware basis in
which {\uefi} initiates a boot sequence and records the hash of the
program to which it passes control by extending a TPM PCR.  Successive
stages of the boot sequence record additional information in the PCRs,
providing a context that controls whether the Attestation Signing Key
\ask{} can be used to sign.  This is a well-established strategy but
it is lengthy, and it intertwines the attestation basis with many
hardware-specific aspects of the boot process.  

Confidential computing provides a more recent---and briefer---route to
software attestation.  In this approach, the early boot phases are not
critical to attestation trustworthiness.  Instead, after the
hardware-specific aspects of boot start a hypervisor, the system
starts a virtual machine within a Trusted Execution Environment.  This
TEE provides protection to the VM within it from the surrounding
untrusted system, and also provides a root of trust for reporting.
Intel's version is called Trusted Domain Extensions (TDX), and AMD has
a Secure Encrypted
Virtualization~\cite{AMD:2020:SEV-SNP-VM-Isolation-Whitepaper}.  We
will focus on AMD's Secure Nested Paging version of SEV.

AMD SEV-SNP uses trusted hardware and firmware (AMD Secure Processor
or AMD-SP) to provide confidentiality and integrity for the runtime
memory of a VM. Confidentiality is provided by encrypting all memory
writes (and thus decrypting all memory reads) performed by the VM,
while integrity is provided by implementing an access control
mechanism that prevents the hypervisor or other VMs from writing to
memory they do not own. It also further subdivides the VM's memory
into up to four privilege levels.  A virtual CPU acting at a higher
privilege level within the VM may grant access to memory pages at
lower privilege levels, choosing a subset of read, write, and execute
permissions.  The page tables for individual vCPUs are an additional
control mechanism, i.e.~an access succeeds only if it resolves under
the page tables and is also authorized for the current privilege
level.

When each VM is launched, SEV-SNP generates a \emph{launch digest} of
the launch detailing configuration options chosen, memory layout, and
a hash of the contents of memory at launch.  The manifest is held by
the Secure Processor AMD-SP, which, when requested by the VM, can
generate an \emph{attestation report} that signs the launch digest
together with a field that contains additional data chosen by the VM.
When the attestation report's additional data contains a public key or
its fingerprint, then the signed manifest effectively certifies that
this VM controls the key.  The key can then be the cryptographic basis
for attestations as required by Maxim~\ref{maxim:display:boot:trait}.

SEV-SNP does not seem amenable to access to a TPM.  In a trust model
where the underlying platform is controlled by a cloud provider, and
may support multiple mutually suspicious tenants, the platform should
not be trusted to ensure TPM communication would always accept
commands from a VM that should be issuing that command, or would
always deliver data (e.g.~secret keys) to a VM that should receive it.
It is uncertain whether SEV-SNP provides another mechanism for
long-term storage of secrets, to be kept and delivered again only when
a new VM has a manifest agreeing with one that executed earlier.
Thus, we regard each launch of a VM as a new principal, and its keys
must be certified anew via the manifest mechanism before other parties
treat it as trusted for a particular purpose.

The upper levels of our CDS (or document control platforms) can be
replicated on SEV-SNP as a hardware root of trust.  However, the
maxims also suggest ways to use the new functionality to improve our
design.

First, Maxim~\ref{maxim:ascend:dependencies} motivates us to eliminate
the cycle by which LKIM depends on the kernel while also measuring it.
The SEV-SNP privilege levels appear to provide several possible
strategies for this.  We will describe one of them here.

LKIM should be able to read the Linux kernel memory, ensuring that a
corrupted kernel cannot lie to LKIM.  However, LKIM should not be able
to modify the kernel memory, which might offer another route to
corrupting the kernel.  Conversely, the Linux kernel should not be
able to modify LKIM's local memory, which could allow it, if
rootkitted, to distort LKIM's report.

If the kernel runs at privilege level 2 and LKIM at the less
privileged level 3, the kernel can have read-write access to its
memory, while LKIM has only read access to this memory.  The memory
allocator at level 0 can use the page tables of the two vCPUs to
ensure that the kernel has no access to LKIM's local memory.  LKIM can
thus reliably determine that the kernel's state respects its intended
invariants, including that IMA remains active with the authorized
policy.  Confidential computing offers this as one solution to the
cyclicity problem of Section~\ref{sec:build:up:lifetime}, restoring
the virtualization-based separations originally envisaged for
LKIM~\cite{Coker::Principles-of-R,thober2008improving}.  The AM and
other ASPs might be located either with the target kernel or with LKIM
depending on details of the scheme.

The TPM locality problem of Section~\ref{sec:build:up:evidence} is
also easier to solve with SEV-SNP.  Our VM may set up a key management
vCPU at privilege level 1---possibly a virtual TPM (vTPM)---ensuring
that its address space is inaccessible to the target kernel and to
LKIM.  If LKIM and the AM communicate with it as a device with
memory-mapped IO, then the normal SELinux mechanisms will allow
controlling access to its command interface.  This vCPU cannot
preserve long-term secrets, but must establish keys each time the VM
is started.  Maxim~\ref{maxim:display:boot:trait} requires us to
certify the new vTPM keys based on the key in the VM manifest.  This
scheme provides a good compliance with
Maxim~\ref{maxim:resist:transitory:corruption} by storing sensitive
keys in areas well-separated from target kernel, which is the
component most accessible to the adversary's potential corruptions.

{Maxim~\ref{maxim:constrain:interaction}} encourages us to separate
the components we need to attest from the remainder of the system, and
running the CDS functionality within a TEE-encapsulated VM provides a
strong hardware-supported separation between these and other system
components.  SELinux is still needed with the CDS VM, because
measuring its policy provides an attestation that the CDS pipeline is
in place and protected from the rest of the VM.  IMA continues to
provide a reliable trust basis to ensure that the CDS executables are
unaffected.  {Maxim~\ref{maxim:short:lived}} yields the same
conclusions as in our original arrangement, because it assures us that
data-driven corruptions cannot change the component behaviors for
subsequent data items.




\section{Related Work}
\label{sec:related}

Remote attestation is first presented by~\citet{Haldar:04:Semantic-Remote} as
introspection of virtual machines.  Our work builds from
\citet{Coker::Principles-of-R} who define the ``Principles of
Remote Attestation'', attestation manager, attestation
protocols, and attestation service providers.
Further extensions of these principles \cite{Banks:2021:Remote-Attestation-Literature-Review,Usman:2023:Attestation-Assurance-Arguments}
also provided insights into how our design and maxims should be structured.
\citet{carpent2018erasmus} introduced the notion of a ``Quality of Attestation
Metric'' that we have not directly utilized for our analysis,
but provides an interesting avenue for future work.

\citet{Rowe:2016bi,Rowe:2016wb} introduces layered attestation and evidence
bundling as a mechanism for building trust from context. Layered attestations
use multiple attestation managers and attestation protocols to assemble
attestations of dependencies into evidence bundles with the attestation target.
The structure of evidence bundles captures semantics of the attestation process.
Bundling and layering are central concepts in the attestation system described here.

Of particular note is the connection between attestation and hardware.
Our work focuses on \emph{hardware-based} attestation where the root of trust is
a hardware component we presume to be uncorruptible. We leverage the TPM and
boot process utilizing techniques presented by \citet{Sailer:04:Design-and-impl}.


Copland~\cite{Ramsdell:2019aa} allows for a diverse set of attestation techniques to be utilized with
a single common interface. In particular, we utilize \emph{binary-based}
attestation techniques~\cite{Brickell:04:Direct-anonymou, Gu:2008:Remote-Attestation-on-Program-Execution,ietf2023rats}
to verify that CDS component binaries are the exact binaries that were expected.
Further, we make use of \emph{property-based} attestations~\cite{Sadeghi:04:Property-based-,Loscocco:07:Linux-kernel-in, Chen:2006:Protocol-Property-Based-Attestation}
as a means for verifying the correctness of components without exact binary
equality being determined. Our utilization of the LKIM tool is a notable
example of this \emph{property-based} attestation technique.


\section{Conclusion}
\label{sec:conc}


Designing effective layered attestation systems is difficult.  One class of challenges are definitional---stating clearly what components and capabilities comprise a target system and its adversary.  In this work we codify a detailed and realistic adversary model for commodity Linux systems equipped with SELinux and IMA.  This adversary model is derived from a top-down security analysis of a Cross Domain Solution application.  The remaining challenges are in choosing and \emph{composing} available attestation and security mechanisms to produce trustworthy attestations of target behavior in the presence of the adversary.  Towards this goal, we motivate a collection of five maxims that guide co-design of a target system and its attestation mechanisms.  The maxims were developed and refined in parallel with adversarial reasoning applied to threats against the CDS example system.  To demonstrate subtleties of the maxims, we walk through a bottom-up design and prototype implementation of the CDS example system that leverages the maxims (when possible) at key decision points in the design.

An empirical analysis of the CDS example system showed that our attestation mechanisms detected realistic attacks derived from our adversary model with negligible performance impact.  Because the maxims guided the design of the CDS system, our positive analysis results support the efficacy of the maxims.  A complementary formal analysis also gives confidence that we didn't overlook any class of attacks within our chosen adversary model.  We also show the broader applicability of the maxims by outlining how they influence the design of two additional example systems with distinct layered architectures.  

Finally, we propose two actionable recommendations for changes to the linux kernel and system architectures to enable future designs that fully adhere to our maxims.  The first is to add virtualizaton support for LKIM to eliminate the cycle between LKIM and the kernel (Recommendation~\ref{recomm:lkim:acyclic}).  The second is to add support for TPM localities in the Linux kernel to fully protect against transient corruptions of the kernel that may lead to permanent misuse of the TPM signing key (Recommendation~\ref{recomm:tpm:localities}).  Realizations of these two recommendations would allow us to strengthen our adversary model and support trustworthy layered attestations for an even wider collection of applications and architectures.

{\footnotesize
  \bibliographystyle{splncsnat
}
  \bibliography{../bib/sldg}

\begin{thebibliography}{49}
\providecommand{\natexlab}[1]{#1}
\providecommand{\url}[1]{\texttt{#1}}
\providecommand{\urlprefix}{}

\bibitem[{{Advanced Micro Devices,
  Inc.}(2020)}]{AMD:2020:SEV-SNP-VM-Isolation-Whitepaper}
{Advanced Micro Devices, Inc.}: {SEV-SNP}: Strengthening vm isolation with
  integrity protection and more (2020),
  \urlprefix\url{https://docs.amd.com/v/u/en-US/SEV-SNP-strengthening-vm-isolation-with-integrity-protection-and-more}

\bibitem[{Banks et~al.(2021)Banks, Kisiel, and
  Korsholm}]{Banks:2021:Remote-Attestation-Literature-Review}
Banks, A.S., Kisiel, M., Korsholm, P.: Remote attestation: A literature review
  (2021), \urlprefix\url{https://arxiv.org/pdf/2105.02466}

\bibitem[{Berger(2025)}]{swtpm}
Berger, S.: {swtpm}: Libtpms-based {TPM} emulator (2025),
  \urlprefix\url{https://github.com/stefanberger/swtpm}

\bibitem[{Boebert and Kain(1985)}]{Boebert1985}
Boebert, W.E., Kain, R.Y.: A practical alternative to hierarchical integrity
  policies.
\newblock In: Proceedings of the Eighth National Computer Security Conference
  (Oct 1985)

\bibitem[{Brickell et~al.(2004)Brickell, Camenisch, and
  Chen}]{Brickell:04:Direct-anonymou}
Brickell, E., Camenisch, J., Chen, L.: Direct anonymous attestation.
\newblock In: Proceedings of the 11th ACM conference on Computer and
  communications security. pp. 132--145. ACM (2004)

\bibitem[{Carpent et~al.(2017)Carpent, ElDefrawy, Rattanavipanon, and
  Tsudik}]{carpent2017lightweight}
Carpent, X., ElDefrawy, K., Rattanavipanon, N., Tsudik, G.: Lightweight swarm
  attestation: A tale of two lisa-s.
\newblock In: Proceedings of the 2017 ACM on Asia Conference on Computer and
  Communications Security. pp. 86--100 (2017)

\bibitem[{Carpent et~al.(2018)Carpent, Tsudik, and
  Rattanavipanon}]{carpent2018erasmus}
Carpent, X., Tsudik, G., Rattanavipanon, N.: Erasmus: Efficient remote
  attestation via self-measurement for unattended settings.
\newblock In: 2018 Design, Automation \& Test in Europe Conference \&
  Exhibition (DATE). pp. 1191--1194. IEEE (2018)

\bibitem[{Castelluccia et~al.(2009)Castelluccia, Francillon, Perito, and
  Soriente}]{Castelluccia:2009:Difficulty-of-Software-Based-Attestation}
Castelluccia, C., Francillon, A., Perito, D., Soriente, C.: On the difficulty
  of software-based attestation of embedded devices.
\newblock In: Proceedings of the 16th ACM Conference on Computer and
  Communications Security. pp. 400--409. CCS '09, Association for Computing
  Machinery, New York, NY, USA (2009),
  \urlprefix\url{https://doi.org/10.1145/1653662.1653711}

\bibitem[{Chen et~al.(2006)Chen, Landfermann, L\"{o}hr, Rohe, Sadeghi, and
  St\"{u}ble}]{Chen:2006:Protocol-Property-Based-Attestation}
Chen, L., Landfermann, R., L\"{o}hr, H., Rohe, M., Sadeghi, A.R., St\"{u}ble,
  C.: A protocol for property-based attestation.
\newblock In: Proceedings of the First ACM Workshop on Scalable Trusted
  Computing. pp. 7--16. STC '06, Association for Computing Machinery, New York,
  NY, USA (2006), \urlprefix\url{https://doi.org/10.1145/1179474.1179479}

\bibitem[{Coker et~al.(2011)Coker, Guttman, Loscocco, Herzog, Millen, O'Hanlon,
  Ramsdell, Segall, Sheehy, and Sniffen}]{Coker::Principles-of-R}
Coker, G., Guttman, J., Loscocco, P., Herzog, A., Millen, J., O'Hanlon, B.,
  Ramsdell, J., Segall, A., Sheehy, J., Sniffen, B.: Principles of remote
  attestation.
\newblock International Journal of Information Security 10(2), 63--81 (June
  2011)

\bibitem[{Costan and Devadas(2016)}]{cryptoeprint:2016:086}
Costan, V., Devadas, S.: Intel {SGX} explained.
\newblock Cryptology ePrint Archive, Report 2016/086 (2016),
  \url{https://eprint.iacr.org/2016/086}

\bibitem[{Eldefrawy et~al.(2012)Eldefrawy, Tsudik, Francillon, and
  Perito}]{ElDefrawy:2012:SMART-Secure-and-Minimal-Architecture-for-Establishing-Dynamic-Root-of-Trust}
Eldefrawy, K., Tsudik, G., Francillon, A., Perito, D.: {SMART:} secure and
  minimal architecture for (establishing dynamic) root of trust.
\newblock In: 19th Annual Network and Distributed System Security Symposium,
  {NDSS} 2012, San Diego, California, USA, February 5-8, 2012. The Internet
  Society (2012),
  \urlprefix\url{https://www.ndss-symposium.org/ndss2012/smart-secure-and-minimal-architecture-establishing-dynamic-root-trust}

\bibitem[{Farmer and andd Vipin~Swarup(1996)}]{FarmerEtAl1996}
Farmer, W.M., andd Vipin~Swarup, J.D.G.: Security for mobile agents: Issues and
  requirements.
\newblock In: Proceedings, 19th National Information Systems Security
  Conference. vol.~2 (1996), \url{https://imps.mcmaster.ca/doc/niss96.pdf}

\bibitem[{Goldman(2019)}]{GoldmanIBMswTPM2019}
Goldman, K.: {IBM} software {TPM} 2.
\newblock GitHub repository (Sept 2019),
  \url{https://github.com/kgoldman/ibmswtpm2/blob/master/src/TcpServerPosix.c#L744}

\bibitem[{Gu et~al.(2008)Gu, Ding, Deng, Xie, and
  Mei}]{Gu:2008:Remote-Attestation-on-Program-Execution}
Gu, L., Ding, X., Deng, R.H., Xie, B., Mei, H.: Remote attestation on program
  execution.
\newblock In: Proceedings of the 3rd ACM Workshop on Scalable Trusted
  Computing. pp. 11--20. STC '08, Association for Computing Machinery, New
  York, NY, USA (2008), \urlprefix\url{https://doi.org/10.1145/1456455.1456458}

\bibitem[{Guttman et~al.(2005)Guttman, Herzog, Ramsdell, and
  Skorupka}]{Guttman:04:Verifying-infor}
Guttman, J.D., Herzog, A.L., Ramsdell, J.D., Skorupka, C.W.: Verifying
  information flow goals in security-enhanced linux.
\newblock Journal of Computer Security 13 (2005)

\bibitem[{Haldar et~al.(2004)Haldar, Chandra, and
  Franz}]{Haldar:04:Semantic-Remote}
Haldar, V., Chandra, D., Franz, M.: Semantic remote attestation -- a virtual
  machine directed approach to trusted computing.
\newblock In: Proceedings of the Third Virtual Machine Research and Technology
  Symposium. San Jose, CA (May 2004)

\bibitem[{Hicks et~al.(2007)Hicks, Rueda, St.Clair, Jaeger, and
  McDaniel}]{Hicks:07:A-logical-speci}
Hicks, B., Rueda, S., St.Clair, L., Jaeger, T., McDaniel, P.: A logical
  specification and analysis for selinux mls policy.
\newblock In: Proceedings of the 12th ACM symposium on Access control models
  and technologies. pp. 91--100. SACMAT '07, ACM, New York, NY, USA (2007),
  \urlprefix\url{http://doi.acm.org/10.1145/1266840.1266854}

\bibitem[{Hutton(2016)}]{hutton2016programming}
Hutton, G.: Programming in haskell.
\newblock Cambridge University Press (2016)

\bibitem[{{IETF RATS Working Group}(2023)}]{ietf2023rats}
{IETF RATS Working Group}: {Remote ATtestation ProcedureS (RATS)}.
\newblock https://datatracker.ietf.org/wg/rats/about/ (2023)

\bibitem[{{Invary Runtime Integrity}(2025)}]{invary_test_kit}
{Invary Runtime Integrity}: {Invary Test Kit}.
\newblock \url{https://github.com/Invary-Runtime-Integrity/invary-test-kit}
  (2025), a Linux kernel module that hijacks the kill system call to test
  Invary's Runtime Integrity sensor software.

\bibitem[{Klabnik and Nichols(2023)}]{klabnik2023rust}
Klabnik, S., Nichols, C.: The Rust programming language.
\newblock No Starch Press (2023)

\bibitem[{Leroy et~al.(2025)Leroy, Doligez, Frisch, Garrigue, R{\'e}my,
  Sivaramakrishnan, and Vouillon}]{leroy25:ocaml}
Leroy, X., Doligez, D., Frisch, A., Garrigue, J., R{\'e}my, D.,
  Sivaramakrishnan, K., Vouillon, J.: The OCaml system release 5.3:
  Documentation and user's manual.
\newblock Inria (2025), \urlprefix\url{https://ocaml.org/manual/5.3/index.html}

\bibitem[{Loscocco and Smalley(2001{\natexlab{a}})}]{Loscocco2001a}
Loscocco, P., Smalley, S.: Integrating flexible support for security policies
  into the linux operating system.
\newblock In: Proceedings of the FREENIX Track: 2001 USENIX Annual Technical
  Conference (FREENIX '01) (Jun 2001{\natexlab{a}})

\bibitem[{Loscocco and Smalley(2001{\natexlab{b}})}]{Loscocco2001b}
Loscocco, P., Smalley, S.: Meeting critical security objectives with
  security-enhanced linux.
\newblock In: Proceedings of the 2001 Ottawa Linux Symposium (Jul
  2001{\natexlab{b}})

\bibitem[{Loscocco et~al.(1998)Loscocco, Smalley, Muckelbauer, Taylor, Turner,
  and Farrell}]{Loscocco:98:The-Inevitabili}
Loscocco, P.A., Smalley, S.D., Muckelbauer, P.A., Taylor, R.C., Turner, S.J.,
  Farrell, J.F.: The inevitability of failure: The flawed assumption of
  security in modern computing environments.
\newblock In: In Proceedings of the 21st National Information Systems Security
  Conference. pp. 303--314 (1998)

\bibitem[{Loscocco et~al.(2007)Loscocco, Wilson, Pendergrass, and
  McDonell}]{Loscocco:07:Linux-kernel-in}
Loscocco, P.A., Wilson, P.W., Pendergrass, J.A., McDonell, C.D.: Linux kernel
  integrity measurement using contextual inspection.
\newblock In: Proceedings of the 2007 ACM workshop on Scalable trusted
  computing. pp. 21--29. STC '07, ACM, New York, NY, USA (2007),
  \urlprefix\url{http://doi.acm.org/10.1145/1314354.1314362}

\bibitem[{Mayer et~al.(2007)Mayer, MacMillan, and
  Caplan}]{Mayer:07:SELinux-by-Exam}
Mayer, F., MacMillan, K., Caplan, D.: {SELinux} by {E}xample.
\newblock Prentice Hall (2007)

\bibitem[{Niu et~al.(2026)Niu, Shi, Han, Liu, Ma, Lyu, and
  Lo}]{Niu:2026:What-You-Trust-is-Insecure}
Niu, Y., Shi, J., Han, R., Liu, Y., Ma, C., Lyu, Y., Lo, D.: What you trust is
  insecure: Demystifying how developers (mis)use trusted execution environments
  in practice (2026), \urlprefix\url{https://arxiv.org/abs/2512.17363}

\bibitem[{Nunes et~al.(2019)Nunes, Eldefrawy, Rattanavipanon, Steiner, and
  Tsudik}]{nunes2019vrased}
Nunes, I.D.O., Eldefrawy, K., Rattanavipanon, N., Steiner, M., Tsudik, G.:
  Vrased: A verified hardware/software co-design for remote attestation.
\newblock In: 28th USENIX Security Symposium (USENIX Security 19). pp.
  1429--1446 (2019)

\bibitem[{Petz and Alexander(2022)}]{petz2022innovations}
Petz, A., Alexander, P.: An infrastructure for faithful execution of remote
  attestation protocols.
\newblock Innovations in Systems and Software Engineering  (2022)

\bibitem[{Petz et~al.(2024)Petz, Thomas, Fritz, Barclay, Schmalz, and
  Alexander}]{Petz:2024:Verified-Configuration-and-Deployment-Paper}
Petz, A., Thomas, W., Fritz, A., Barclay, T.J., Schmalz, L., Alexander, P.:
  Verified configuration and deployment of layered attestation managers.
\newblock In: Software Engineering and Formal Methods: 22nd International
  Conference, SEFM 2024, Aveiro, Portugal, November 6-8, 2024, Proceedings. pp.
  290--308. Springer-Verlag, Berlin, Heidelberg (2024)

\bibitem[{{Phoronix Media}(2026)}]{phoronix_test_suite}
{Phoronix Media}: Phoronix test suite.
\newblock \url{https://www.phoronix-test-suite.com/} (2026), gitHub repository:
  \url{https://github.com/phoronix-test-suite/phoronix-test-suite}

\bibitem[{Ramsdell et~al.(2019)Ramsdell, Rowe, Alexander, Helble, Loscocco,
  Pendergrass, and Petz}]{Ramsdell:2019aa}
Ramsdell, J., Rowe, P.D., Alexander, P., Helble, S., Loscocco, P., Pendergrass,
  J.A., Petz, A.: Orchestrating layered attestations.
\newblock In: Principles of Security and Trust (POST'19). Prague, Czech
  Republic (April 8-11 2019)

\bibitem[{Ramsdell(2020)}]{Ramsdell:2020:Chase}
Ramsdell, J.: Chase: A model finder for finitary geometric logic.
\newblock \url{https://github.com/ramsdell/chase} (2020)

\bibitem[{Rowe(2016{\natexlab{a}})}]{Rowe:2016wb}
Rowe, P.D.: {Confining adversary actions via measurement}.
\newblock Third International Workshop on Graphical Models for Security pp.
  150--166 (2016{\natexlab{a}})

\bibitem[{Rowe et~al.(2021)Rowe, Ramsdell, and Kretz}]{Rowe-Automated-Trust-21}
Rowe, P., Ramsdell, J., Kretz, I.: Automated trust analysis of copland
  specifications for layered attestations.
\newblock In: Principles and Practice of Declarative Programming (PPDP 21) (Sep
  2021)

\bibitem[{Rowe(2016{\natexlab{b}})}]{Rowe:2016bi}
Rowe, P.D.: {Bundling Evidence for Layered Attestation}.
\newblock In: Trust and Trustworthy Computing, pp. 119--139. Springer
  International Publishing, Cham (Aug 2016{\natexlab{b}})

\bibitem[{Sadeghi and St{\"u}ble(2004)}]{Sadeghi:04:Property-based-}
Sadeghi, A., St{\"u}ble, C.: Property-based attestation for computing
  platforms: caring about properties, not mechanisms.
\newblock In: Proceedings of the 2004 workshop on New security paradigms. pp.
  67--77. ACM (2004)

\bibitem[{Sailer et~al.(2004{\natexlab{a}})Sailer, Zhang, Jaeger, and van
  Doorn}]{Sailer:04:Design-and-impl}
Sailer, R., Zhang, X., Jaeger, T., van Doorn, L.: Design and implementatation
  of a tcg-based integrity measurement architecture.
\newblock In: Proceedings of the 13th USENIX Security Symposium. USENIX
  Association, Berkeley, CA (2004{\natexlab{a}})

\bibitem[{Sailer et~al.(2004{\natexlab{b}})Sailer, Zhang, Jaeger, and
  Doorn}]{sailer2004design}
Sailer, R., Zhang, X., Jaeger, T., Doorn, L.V.: Design and implementation of a
  tcg-based integrity measurement architecture.
\newblock In: USENIX Security symposium. vol.~13, pp. 223--238
  (2004{\natexlab{b}})

\bibitem[{Shacham(2007)}]{Shacham:2007:Geometry-of-Innocent-Flesh-on-the-Bone}
Shacham, H.: The geometry of innocent flesh on the bone: return-into-libc
  without function calls (on the x86).
\newblock In: {ACM} Conference on Computer and Communications Security {CCS}.
  pp. 552--561. {ACM} (2007),
  \urlprefix\url{https://doi.org/10.1145/1315245.1315313}

\bibitem[{Sultana et~al.(2022)Sultana, Shands, and
  Yegneswaran}]{sultanasy22acase}
Sultana, N., Shands, D., Yegneswaran, V.: A case for remote attestation in
  programmable dataplanes.
\newblock In: Proceedings of the 21st {ACM} Workshop on Hot Topics in Networks,
  HotNets 2022, Austin, Texas, November 14-15, 2022. pp. 122--129. {ACM}
  (2022), \urlprefix\url{https://doi.org/10.1145/3563766.3564100}

\bibitem[{Thober et~al.(2008)Thober, Pendergrass, and
  McDonell}]{thober2008improving}
Thober, M., Pendergrass, J.A., McDonell, C.D.: Improving coherency of runtime
  integrity measurement.
\newblock In: {ACM} workshop on {Scalable Trusted Computing}. pp. 51--60 (2008)

\bibitem[{{Trusted Computing Group}(2019)}]{TCGReservedLocalities2019}
{Trusted Computing Group}: Registry of Reserved TPM 2.0 Handles and Localities,
  1.00 edn. (February 2019),
  \urlprefix\url{https://trustedcomputinggroup.org/wp-content/uploads/RegistryOfReservedTPM2HandlesAndLocalities_v1p1_pub.pdf}

\bibitem[{{Trusted Computing
  Group}(2025)}]{Trusted-Computing-Group:2025:TCG-TPM-2.0-Part0-Introduction}
{Trusted Computing Group}: Trusted Platform Module 2.0 Library Part 0:
  Introduction.
\newblock Trusted Computing Group, 3855 SW 153rd Drive, Beaverton, OR 97003,
  version 184 edn. (March 2025),
  \urlprefix\url{https://trustedcomputinggroup.org/wp-content/uploads/Trusted-Platform-Module-2.0-Library-Part-0-Version-184_pub.pdf}

\bibitem[{Usman et~al.(2023)Usman, Cole, Asplund, Boeira, and
  Vestlund}]{Usman:2023:Attestation-Assurance-Arguments}
Usman, A.B., Cole, N., Asplund, M., Boeira, F., Vestlund, C.: Remote
  attestation assurance arguments for trusted execution environments.
\newblock In: Proceedings of the 2023 ACM Workshop on Secure and Trustworthy
  Cyber-Physical Systems. pp. 33--42. SaT-CPS '23, Association for Computing
  Machinery, New York, NY, USA (2023),
  \urlprefix\url{https://doi.org/10.1145/3579988.3585056}

\bibitem[{Wikipedia(Accessed {Jan.}~2026)}]{WikipediaTurtlesalltheway}
Wikipedia: Turtles all the way down.
\newblock \url{https://en.wikipedia.org/wiki/Turtles_all_the_way_down}
  (Accessed {Jan}~2026)

\bibitem[{Xu et~al.(2023)Xu, Lu, Du, Ding, Li, Wu, Payer, and
  Mao}]{xu2023silent}
Xu, J., Lu, K., Du, Z., Ding, Z., Li, L., Wu, Q., Payer, M., Mao, B.: Silent
  bugs matter: A study of compiler-introduced security bugs.
\newblock In: USENIX Security Symposium. pp. 3655--3672 (2023)

\end{thebibliography}
  %
}

\section{Additional system detail}
\label{sec:appendix:detail}
\subsection{The Attestation Manager Subsystem}
\label{sec:build:up:am}

A key and novel aspect of our layered attestation approach is the "hand-off" between boot-time and run-time attestation.  We call the collection of mechanisms that facilitate this transition the \emph{Attestation Manager Subsystem}.  Among these is a server process called the \emph{Attestation Manager Launcher} that listens for (run-time) requests and launches an \emph{Attestation Manager} AM instance, which in turn orchestrates auxiliary executables called \emph{Attestation Service Providers} or ASPs.  These components operate in userspace, but \emph{leverage system-level mechanisms} both for self-protection and as a providers of evidence from deeper layers in the architecture -- hence the hand-off.  While this architecture supports more flexible and semantically-rich attestations, care must be taken to justify its robustness and conformance to the Design for Attestation maxims introduced in Section~\ref{sec:downward}.

Towards these goals, we instantiate the Attestation
Manager using the {\maestro}\footnote{Measurement and
  Attestation Execution and Synthesis Toolkit for Remote
  Orchestration.} toolkit for layered attestation
systems~\cite{Petz:2024:Verified-Configuration-and-Deployment-Paper}.
{\maestro} provides tools for specifying Copland~\cite{Ramsdell:2019aa} attestation protocols, then synthesizing Attestation Manager configurations and executable procedures that gather, bundle, and appraise structured evidence.  
{\maestro} AMs support sound and sufficient environments for Copland
attestation protocol execution:

\begin{description}
  \item [\textsc{Configuration Sufficiency:}] The static configuration of a {\maestro}-configured AM suffices to execute a Copland attestation protocol without runtime errors due to missing primitives or configuration.
  \item [\textsc{Execution Soundness:}] Protocol execution within a
  {\maestro}-configured AM is guaranteed correct w.r.t. the
  Copland reference semantics -- Genuine measurement (appraisal) primitives occur in a precise order and evidence is bundled (unbundled)
  reliably, ensuring Copland evidence appraisals accurately reflect the
  measured system's state.
\end{description}

Notice that \textsc{Configuration Sufficiency} is an \emph{availability} property:  If attestation is a game to disclose specific corruption attempts by the adversary, it ensures that we can at least \emph{play the game we want to play}.  \textsc{Execution Soundness} is what leads us to believe we have legitimately won the game when the AM Subsystem tells us we did.  On target systems with rich system-level security mechanisms, \textsc{Configuration Sufficiency} provides more value than \textsc{Execution Soundness} -- soundness can be delegated to the trusted mechanisms.  Regardless of target system, \textsc{Execution Soundness} is critical in the context of Copland \emph{evidence appraisal}.  When a client recieves a response from the AM Launcher, it must have a trusted procedure to unbundle Copland evidence and make semantic judgements about the target system.  Formal verification in the Rocq
proof assistant of {\maestro}-synthesized components gives us high confidence in the soundness of this process \cite{petz2022innovations}.

Having robust AM component implementations is not enough.  Their integration into target systems must also be practical and support the Design for Attestation maxims:

\noindent\textbf{Maxim~\ref{maxim:constrain:interaction} (constrained interaction)}
Access to the AM Subsystem is managed by SELinux policy:  the only external-facing interface is the AM Launcher, and within the AM Subsystem the core {\maestro} AM has exclusive access to a predictable set of ASP executables.  The {\maestro} AM functionality is minimal by design and formally verified.  While the ASP executables may be arbitrarily complex, they must abide by their system-enforced SELinux boundaries.    

\noindent\textbf{Maxim~\ref{maxim:short:lived} (short-lived processes)}
The AM Launcher never reads input; for each new connection it spawns a fresh (and short-lived) MAESTRO AM instance.  The latter, even if corrupted by a malicious crafted input, will never read a subsequent input and thus will never answer a genuine query erroneously.  

\noindent\textbf{Maxim~\ref{maxim:resist:transitory:corruption} (transitory corruption)}
The AM Subsystem never stores secrets or directly endorses attestation results.  Rather, it orchestrates invocation of measurement ASPs and ultimately invokes a signing ASP to endorse evidence on its behalf.

\noindent\textbf{Maxim~\ref{maxim:display:boot:trait} (evidence displays boot traits)}
The AM, AM Launcher, and ASP executables are all in predictable locations known to system-level policies, and thus can be reflected in evidence to bolster the integrity of the attestation process itself.

\noindent\textbf{Maxim~\ref{maxim:ascend:dependencies} (acyclic dependencies)}
The AM Subsystem may \emph{collect} evidence produced at lower layers of the system (i.e. boot-time or kernel-level measurements), but the AM runs as a normal user-level process.  Measurements of the AM Subsystem itself are managed and protected by system-level mechanisms, and thus do not introduce acyclic measurement dependencies.

\subsection{Requirements on the SELinux and IMA policies}
\label{sec:policy:reqs}

Our system relies on IMA and SELinux to protect attestation infrastructure
and \ask{}.

The IMA policy appraises all attestation infrastructure and CDS binaries and configurations.
We accomplish this by running these under a particular
user and appraising all executables owned or executed by that user. An IMA signature is used
on these files rather than just a hash to ensure an attacker rebooting the machine
into an uncontrolled state cannot modify the attestation infrastructure and then
simply rehash it. A broader policy, such as appraising all executables owned by
all users, would provide stronger protections for the whole system but is not
required for proper functioning of our attestation mechanisms.

SELinux policy is used to protect both the attestation framework and the CDS.
The policy for the attestation framework
(1) prevents \ask{} and its delegated key usage policy from being read
by anyone other than the \texttt{tpm\_sign} ASP,
(2) ensures only the AM can run the ASPs,
(3) restricts the ASPs to only the permissions needed for their job,
and (4) protects the attestation framework from interference from other parts of the system.
The policy for the CDS
(1) enforces an assured pipeline ~\cite{Boebert1985} on the processing of emails in the system,
and (2) protects the CDS from other parts of the system.

Giving unique SELinux labels to the executables, processes, files, and ports used by the attestation
framework and CDS ensures that SELinux will not allow the
other parts of the system to interfere with them thus fulfilling
\autoref{maxim:constrain:interaction}.
The policy for the assured pipeline uses unique
SELinux labels for all of the executables, processes, and directories involved.
Each stage can only read from the previous directory in the pipeline and write
to the next directory. In addition, each stage can only be started by the
previous stage. Thus the flow in Figure~\ref{fig:cds} is enforced. 



{\footnotesize
  \ifacm\else\ifllncs{}
    \else
      {\newpage\footnotesize\tableofcontents}
    \fi
  \fi
}

\end{document}